\RequirePackage{snapshot}
 \documentclass[annual]{acmsiggraph}

\usepackage{smrdefaults}
\usepackage[usenames,dvipsnames]{color}
\usepackage{array}
\usepackage{subfig}
\usepackage{multirow}
\usepackage[title]{appendix}

\TOGonlineid{0000}

\TOGvolume{}
\TOGnumber{}

\TOGarticleDOI{}

\TOGprojectURL{}
\TOGvideoURL{}
\TOGdataURL{}
\TOGcodeURL{}

\title{Interactive 3D Face Stylization Using Sculptural Abstraction}

\author{
  Jan Jachnik$^{1}$ $\qquad\qquad$ Dan B Goldman$^{2}$
  $\qquad\qquad$ Linjie Luo$^{2}$ $\qquad\qquad$ Andrew J. Davison$^{1}$\\ \vspace{-5pt}\\
  $^1$Imperial College London\\ 
  $^2$Adobe Research}

\pdfauthor{Jan Jachnik, Dan B Goldman, Linjie Luo, Andrew J. Davison}

\newcommand{\eg}{\emph{e.g.{}}}
\newcommand{\etal}{{\em et~al.}}

\newcommand{\vect}[1]{\mathbf{#1}}

\begin{document}


\teaser{  
\def\bust{teaser} 
\def\imw{0.19\linewidth} 
\subfloat[Input mesh]{
	\label{teaser:input}
	\includegraphics[width=\imw]{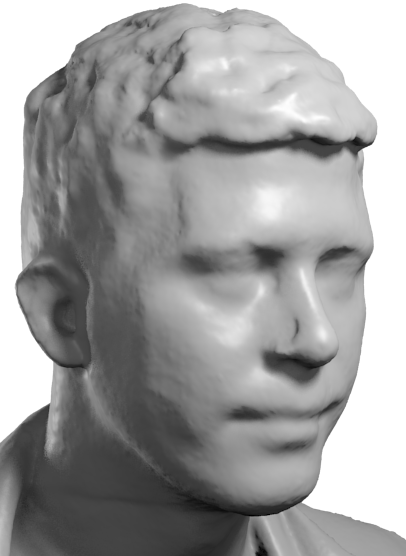}
}
\subfloat[Input segmented]{
	\label{teaser:segmented}
	\includegraphics[width=\imw]{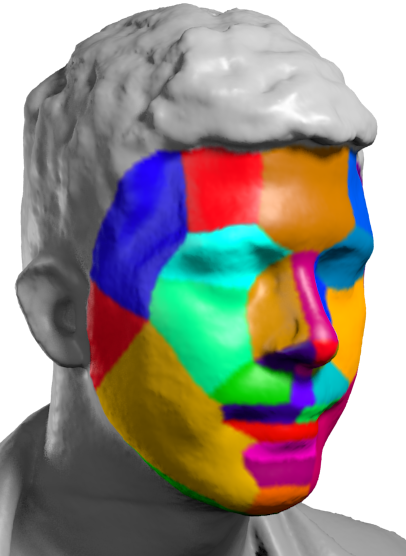}
}
\subfloat[Abstracted mesh]{
	\label{teaser:coarse}
	\includegraphics[width=\imw]{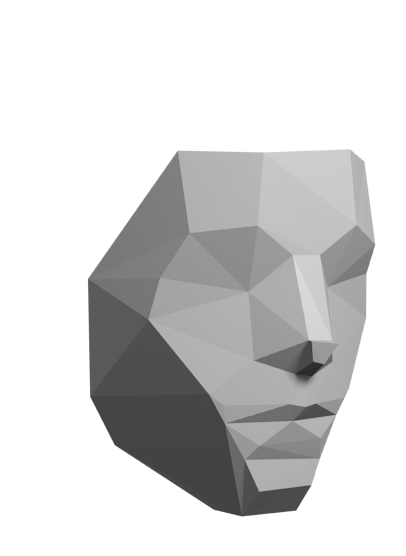}
}
\subfloat[Deformed abstracted mesh]{
	\label{teaser:deformed}
	\includegraphics[width=\imw]{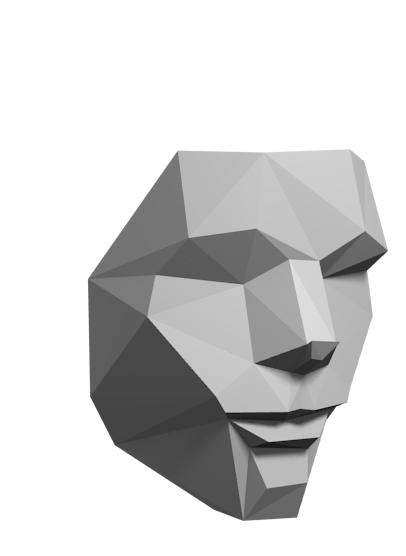}
}
\subfloat[Output mesh]{
	\label{teaser:abstracted}
	\includegraphics[width=\imw]{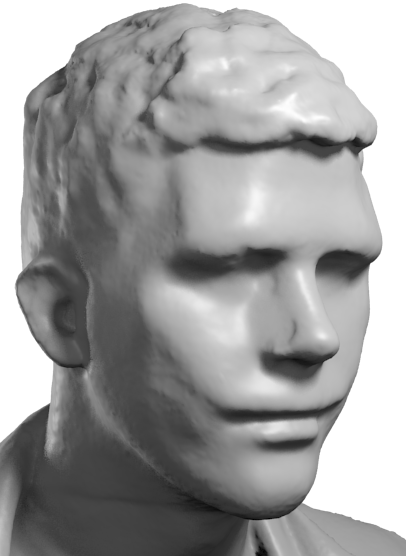}
}
  \caption{
We transform a scanned input mesh into a stylized output emphasizing anatomy and characteristic features. The input is segmented into nearly planar regions via alignment with a template model. The angular definition of the segmented mesh is enhanced in a highly controllable optimization which generates smooth full resolution output interactively.
\label{fig:teaser}
}
}

\maketitle


\begin{abstract}
Sculptors often deviate from geometric accuracy in order to enhance the appearance of their sculpture. These subtle stylizations may emphasize anatomy, draw the viewer's focus to characteristic features of the subject, or symbolize textures that might not be accurately reproduced in a particular sculptural medium, while still retaining fidelity to the unique proportions of an individual. In this work we demonstrate an interactive system for enhancing face geometry using a class of stylizations based on visual decomposition into abstract semantic regions, which we call \emph{sculptural abstraction}. We propose an interactive two-scale optimization framework for stylization based on sculptural abstraction, allowing real-time adjustment of both global and local parameters. We demonstrate this system's effectiveness in enhancing physical 3D prints of scans from various sources.
\end{abstract} 

\section{Introduction}
\label{sec:intro}
The technology required for a ``3D fax'' that can scan the shapes of household objects of various sizes, transmit them electronically, and reprint facsimiles of those objects at a distant location, is now available at a price feasible for many households. In addition to utilitarian uses such as printing missing or broken parts for household objects, 3D scanning and printing tools are also being employed for  more personal items such as figurines of family and friends\footnote{\eg, \url{www.shapify.me}, \url{www.shapeways.com/tutorials/shapeme}, \url{www.fablitec.com}}. Although the market for such tools is nascent, we suggest that given the appropriate tools and low enough pricing in the future, we might be just as likely to send our relatives a 3D printed figure of a family vacation as we would a framed photograph: Just as photography democratized 2D imagery by unchaining it from the painter's hand, 3D scanning and printing may democratize 3D representations by decoupling them from the sculptor's hand.

Thus far, research on 3D scanning and printing has rightly focused on geometric accuracy. However, 3D printed human figures often appear lifeless, particularly when generated with commodity scanning and printing. We propose that creative applications of these technologies require further exploration of \emph{sculptural stylization}, which we define to be when a 3D model deviates from geometric accuracy in a manner analogous in many ways to that of non-photorealistic rendering for 2D imagery. Sculptural stylization could be
either for purely aesthetic purposes; or to improve the recognizability of the subject by emphasizing characteristic features~\cite{Winnemoller:RVA2006}; or to fit the limitations or requirements of a given sculpting medium; or simply to imitate historically popular styles of sculpture.  
Well known sculptors consistently exaggerate away from geometric accuracy in order to achieve some combination of these aims (a preliminary analysis of how sculpted busts deviate from real human geometry is provided in Section \ref{sec:analysis}).

As described in subsequent sections, one important class of stylizations first decomposes a model into masses, planes or other components. These decompositions could be purely geometric, but are often based on higher-level object semantics or a knowledge of underlying anatomy. We use the term \emph{sculptural abstractions} to describe these semantically-based decompositions, using the word \emph{abstraction} in the sense of isolating essential qualities.

In this paper we propose an interactive approach to stylization of human faces, founded on a specific sculptural abstraction used by artists called
\emph{the planes of the head} --- regions of the surface of the face which, although not literally planar, may be grouped
together by geometric consistency or separated by underlying anatomy. We generalize this concept to scans of other
objects beyond simply faces, and refer to the general concept as \emph{sculptor's planes}. Our method's stylizations emphasize the
contrast between these forms by accentuating the angles between them and making the regions within them more planar. We imagine an expert sculptor who first studies his subject
and identifies these sculptor's planes, then produces a rough, low-detail model in which the surface angles between
planes are exaggerated and deviations within planes are smoothed, and finally introduces fine details of the surface.

Our system, using a 3D scan of the subject as its input, supports an amateur digital sculptor in several stages: First, the
sculptor's planes are automatically identified --- either by aligning a scan to a generic, pre-segmented head model, or by automatic geometry analysis; and the scan is then simplified to an abstracted mesh using the given segmentation. Second, we
construct and optimize an energy function to balance the exaggeration of angles of this abstracted mesh with preservation of the original form and some key individual facial characteristics. The sculptor can adjust both the global scale and local scale of exaggeration as well as fidelity to the facial characteristics. Finally,
we map the resulting deformation back to the original full resolution scan, and the sculptor can locally adjust the the amount of detail to include from the original scan, as well as the smoothness between sculptors planes. The last two stages are performed continuously at interactive frame rates, providing detailed user control of both global and local parameters.

Our contributions include an algorithmic analysis of stylization based on sculptural abstraction, and an interactive system to emulate these techniques under real-time user control of intuitive parameters. We show that our approach permits a wide array of both subtle and exaggerated stylizations on both faces and other models, and illustrate qualitative improvement over previous geometry deformations applied to faces.

\section{Related Work}
\label{sec:related}
The non-photorealistic rendering literature features a long history of transforming photographs and 3D models into artistic depictions. In particular, Gao~\etal~\shortcite{Gao:etal:2013} used an abstracted model similar to the planes of the head for 2D stylization and exaggeration. Winnemoller~\etal~\shortcite{Winnemoller:RVA2006} demonstrated that certain types of abstraction improve human memory tasks, including facial recognition and scene recall. We hypothesize that sculptural abstraction can play a similar role for 3D shape. Furthermore, many papers in NPR explore the depiction of 3D scenes to improve shape understanding. Hertzmann and Zorin developed a hatching technique to emphasize aspects of curved surfaces that are otherwise hard to perceive~\cite{Hertzmann:2000:ISS}. Cole~\etal~\shortcite{Cole:2008:WDP} analyzed where people draw lines when representing 3D surfaces, which might be similar to the features that sculptors exaggerate.

We are unaware of previous work that defines the broader notion of \textit{sculptural abstraction} in 3D geometry. However, the idea of 3D collage~\cite{Gal:etal:NPAR2007} seems to fall within this scope, as it attempts to represent the gestalt of a given 3D shape by aggregating several smaller 3D primitives. Bhat~\etal~\shortcite{Bhat:etal:SGP2004} also applied a method similar to image analogies~\cite{Hertzmann:etal:SIGGRAPH2001} in the context of 3D geometry. This method applied a learned geometric texture to an entire model rather than local deformations based on existing geometry, and thus could be used as a complement to our sculptural abstraction in order to reproduce the surface texture imposed by various real-world sculpting tools.

Other existing methods for automatic non-metric mesh enhancement take approaches which have predictable and global effects.
Blanz and Vetter \shortcite{Blanz:Vetter:CGIT1999} fitted a parametric 3D model built from a large family of faces to a new scan, and could create caricatures by shifting the parameters away from the mean face. This results in a global linear deformation, in contrast to our nonlinear template-driven deformation. PriMo~\cite{Botsch:2006:PCP} is another deformation strategy that has been used to demonstrate caricature, by modeling polygons as prisms of finite thickness and constructing an energy function to increase angles between them. This resembles our exaggeration energy term, but although our system can be used for stylizations such as caricature --- \eg exaggeration of key individual features --- we characterize our goal of abstraction as sharpening visual contrast and form \emph{without} modifying key individual features. This distinction is explored in detail in Section~\ref{sec:Lanteri}, in which we propose \emph{Lanteri constraints} for maintaining fidelity to these features.

Various signal processing techniques exist for meshes which can exaggerate features. The simplest is the 3D equivalent of the unsharp mask, where a 3D shape is enhanced  by smoothing its normals and then exaggerating their angles in the opposite direction~\cite{Yagou:etal:3DI2003}. Eigensatz~\etal~\shortcite{Eigensatz:etal:2008} operate in the curvature domain, but do not process meshes in real time.
More recently, anisotropic smoothing and exaggeration have been implemented in graphics hardware~\cite{Chuang:Kazhdan:SIG2011}. Although these methods can produce some results similar to ours, they don't offer simple local control.

Some previous works have explored automated abstraction of shapes~\cite{Mehra:etal:2009} and shape collections~\cite{Yumer:etal:2012} into semantic groupings, though they do not demonstrate stylizations based on these groupings. Similarly, Lee~\etal~\shortcite{Lee:etal:2000} decomposed geometry into a base mesh plus displacements not unlike our two-scale decomposition, but their model is not suitable for deformation because it associates each high-resolution vertex with only one base polygon.

\section{Background}
\label{sec:background}
\begin{figure}[t]
	\begin{center}
		\includegraphics[width=0.9\columnwidth]{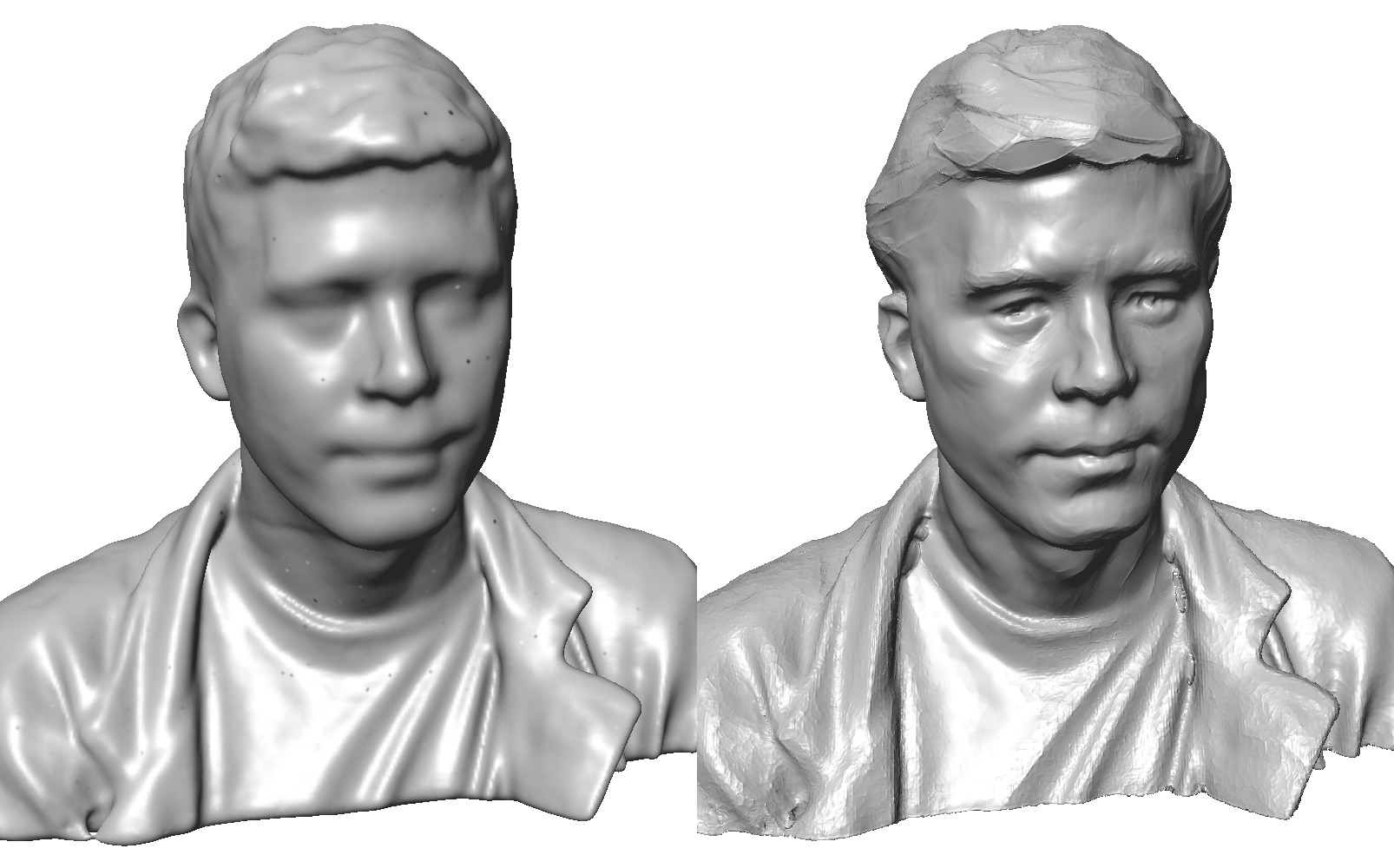}
	\end{center}
	\caption{\label{fig:gio}
		Left: A scan acquired using KinectFusion. Right: The same scan after sculpting by Gio Nakpil. Note exaggerated features such as cheekbones, temples, jawline, chin, and sides of the nose.
	}
\end{figure}

In order to learn what techniques sculptors use in their abstraction, we interviewed two accomplished professionals: Gio Nakpil, a sculptor who works both in physical and digital sculpture, including work for Industrial Light and Magic and Valve; and Mike Magrath, an instructor at Seattle's Gage Academy of Art. Although many techniques are particular to specific styles, we narrowed our focus to abstraction techniques for human busts that span multiple styles.

Two common points of reference between the sculptors emerged: First, both sculptors mentally and visually decompose a subject into component \emph{masses} or \emph{forms} at a variety of scales --- starting from larger groupings such as the forehead, and working down to smaller details like the sides of the nose and the folds of the eyelid. And second, in the case of human faces this decomposition is often described in terms of \emph{planes of the head}, as defined in Section~\ref{sec:intro}.

Nakpil emphasized the role of highlight and shadow in sculpting technique\footnote{In this context, highlight and shadow refer to the broad artistic definition of regions brighter or darker than the mean, not only specular highlights and cast shadows.}, noting that sculptors view their work under a range of lighting conditions and from many angles to understand and control the shapes and relationships of those highlights and shadows. Since sculptures are typically lit from above, sculptors may tilt surfaces toward the horizontal, exaggerating the contrast between highlights and shadows. Figure~\ref{fig:gio} shows a scan we provided to Nakpil alongside his sculptural interpretation using ZBrush.

Magrath made special note of the exaggeration of ``soft'' and ``hard'' forms. This nomenclature refers to the distinction between angular regions with rapidly alternating high and low curvature (\eg, the bridge of a nose) versus rounded, uniform medium-curvature regions (\eg, the forehead). He characterized his abstraction technique as ``making the hard forms harder, and the soft forms softer.''

The notion of \emph{planes of the head} permeates literature about sculpting, and dates back at least to the turn of the 19th century, when the sculptor Edouard Lanteri (whom Rodin described as ``Dear Master''~\cite{Lanteri:1911}) wrote a seminal textbook. Although much of his work describes technical methods for ensuring metric accuracy, he also suggests a methodical approach to exaggerating these planes and their junctions using side lighting to reveal them more clearly to the sculptor:

\begin{quote}
	When these divisions of form have been obtained in their proper drawing by studying each separately, the work may appear a little hard. Then it becomes necessary to work by colour --- that is to say, by the comparative values of the half-tints, in simplifying or accentuating the surfaces or planes which divide these forms.~\cite{Lanteri:1985}
\end{quote}

\begin{figure}[t]
	\begin{center}
		\includegraphics[height=0.6\columnwidth]{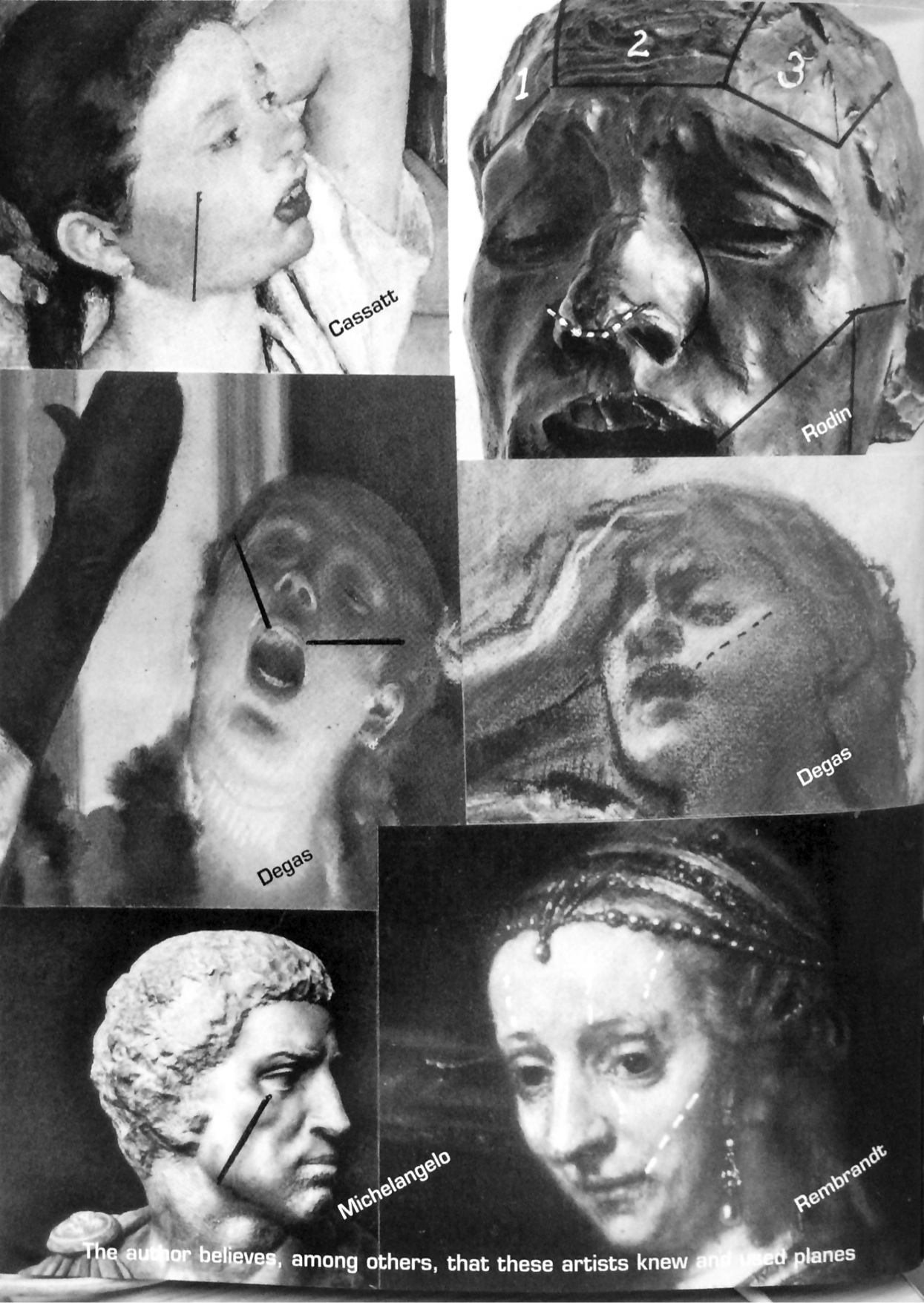}
		\includegraphics[height=0.6\columnwidth]{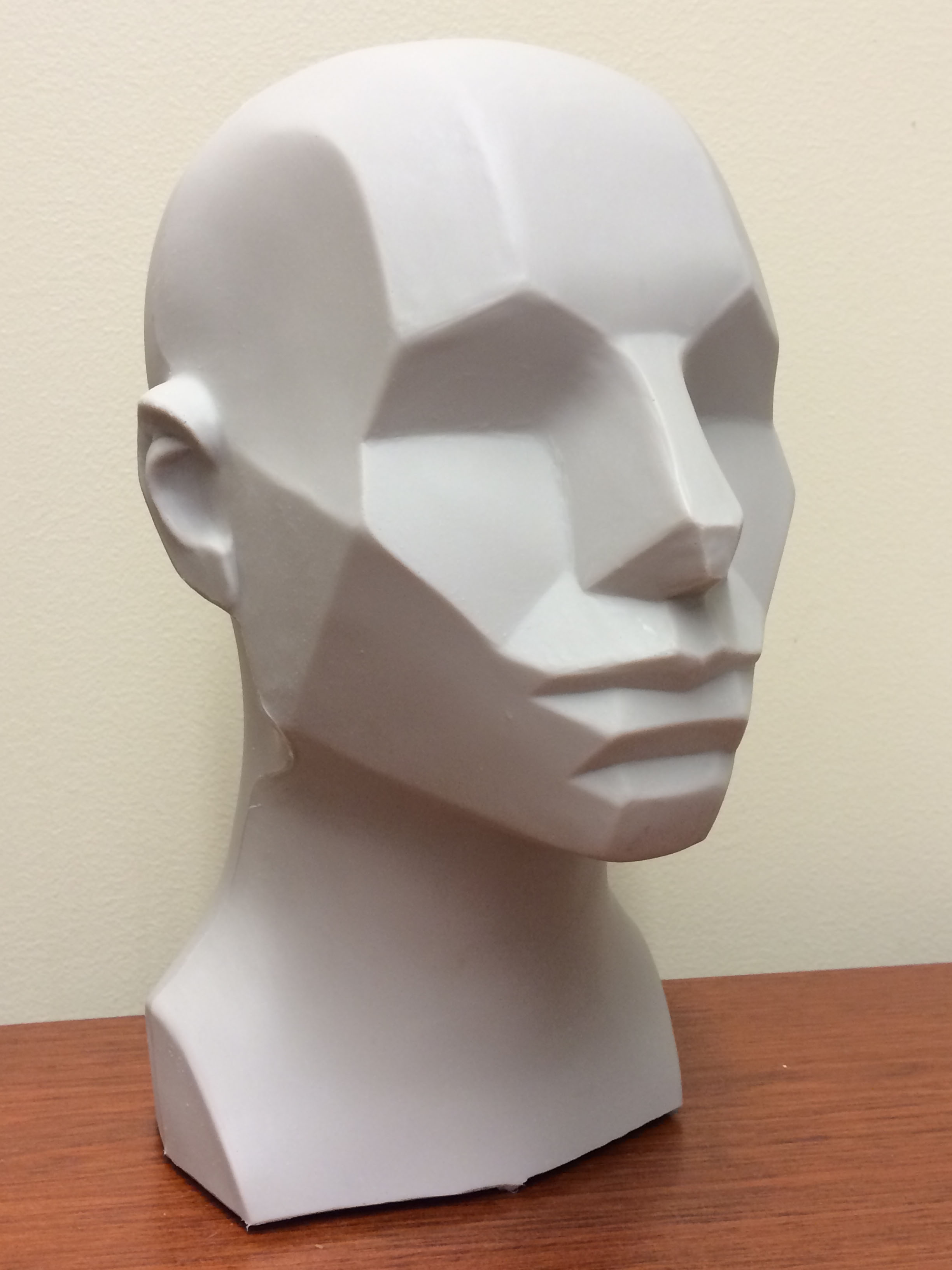}
	\end{center}
	\caption{\label{fig:asaro}
		Left: Examples of planar decompositions in painting and sculpture, \copyright John Asaro 1976, excerpted with permission.  Right: A variation of the \emph{Planes of the Head} entitled \emph{Memorized}, intended for student memorization.
	}
\end{figure}

The concept of \emph{planes of the head} was further formalized in the late 20th century, most notably by John Asaro, an instructor at
Art Center College in Pasadena, California~\cite{planesofthehead}. Asaro hypothesized that decompositions of facial form into masses and
planes have been used by painters and sculptors since antiquity, and proposed planar decompositions for works by Cassatt,
Rodin, Degas, Rubens, Vermeer, Michelangelo and many others. As an aid to students of sculpting and painting, Asaro
developed a set of canonical busts which exaggerate the planes to their logical extreme, reducing a human head to a nearly polygonal
object (see Figure~\ref{fig:asaro}).

While sculptors use these planes to exaggerate certain forms, Lanteri's writing also makes it clear that certain aspects of human anatomy and individual facial characteristics are perceptually relevant, and must not be perturbed by these stylizations. He writes voluminously about using calipers to ensure that certain distances, ratios, and angles are preserved in a sculpture of a given model. Here is one such passage:

\begin{quote}
	From the tip of the nose measure the distance to the most projecting part of the chin, i.e. the subcutaneous mental eminence... To make sure of the correctness of the previous measure, you now take the distance from the ear to the mental eminence (see Fig. 40) in your calipers, and if this measure, taken from both ears of course, coincides with the previously fixed point on the chin, your measure runs a fair chance of being correct. If it is not, you have to... retake both measures, until they agree.~\cite{Lanteri:1985}
\end{quote}
Additional measurements proposed by Lanteri appear in Appendix \ref{app:Lanteri}. We make use of these in our algorithm to preserve the facial characteristics of the subjects during stylization.

We hypothesize that the \emph{planes of the head} perform a similar function to lines in 2D illustration: Artists abstract detail that isn't important and emphasize features that do matter for recognition and understandability. Although we may not yet know exactly what features are perceived as important for emphasis in 3D, we choose to follow methods employed by professional sculptors because it is reasonable to assume they relate to perceptual importance.

Thus the challenge of sculptural abstraction is to enhance or exaggerate visual contrast between sculptor's planes, while simultaneously preserving certain critical global geometric relationships. Although it may seem these goals are fundamentally incompatible, in the sections that follow we demonstrate a system that accomplishes this automatically, interactively and under flexible user control.

\subsection{Analysis of scans of humans vs. sculptures}
\label{sec:analysis}
We observed informally that, in spite of some sculptors' focus on geometric fidelity, some features in realistic classical and modern sculptures appear exaggerated. This is hard to detect when comparing people directly to sculptures, because their reflectance qualities differ so drastically. However, these differences become more apparent when people and sculptures are scanned. Perhaps most critically for our application, 3D prints of scanned faces often appear flat and lifeless.

An ideal analysis of sculptural stylization would require extensive data collection of existing sculptures and comparisons to human scans of the individual models being sculpted. In the absence of such a dataset, we performed an informal comparison of the depth of the eye sockets between six scans of humans -- four from KinectFusion~[Newcombe et al.~2011] and two using the method of Beeler~\etal~[2010] -- and five scans of sculptures -- three of Rodin's ``Burghers of Calais''\footnote{\url{http://www.stanford.edu/~qianyizh/projects/scenedata.html}}, and two busts from the MIT CSAIL 3D Model Database\footnote{\url{http://people.csail.mit.edu/tmertens/textransfer/data/}}. First, a number of feature points on the models were identified and manually labeled. Then, we measured the depth of the eye sockets using four separate scale-invariant measures: Measure A takes the distance from the midpoint of the inner corners of the eye to the plane formed by the midpoint of each eyebrow and the tip of the chin, normalized using the distance between the points at the base of the ears. Measure B is the same as measure A, but using the outer corners of the eyes. Measure C takes the distance from the midpoint of the inner corners of the eye to the plane formed by the saddle point at the bridge of the nose and the corners of the mouth, also normalized using the distance between the ears. Measure D is the same as C, but using the outer corners of the eyes. Each measure was aggregated across human and sculpted scans using both the mean and the median. A summary of the results can be found in Table~\ref{tab:humvsculpt}.

\begin{table}[htb]
\begin{center}
	\begin{tabular}{ l l | c | c | c | c | }
		\cline{3-6}
		 & & A & B & C & D \\
		\cline{1-6}
		\multicolumn{1}{ |c| }{\multirow{3}{*}{Mean}}
			& Human & 0.069 & 0.134 & 0.094 & 0.157 \\
		\cline{2-6}
		\multicolumn{1}{ |c|  }{}
			& Sculpt & 0.112 & 0.177 & 0.131 & 0.193 \\
		\cline{2-6}
		\multicolumn{1}{ |c|  }{}
			& \textbf{\% Increase} & \textbf{63.1} & \textbf{32.5} & \textbf{39.3} & \textbf{23.1} \\ \hline
		\multicolumn{1}{ |c| }{\multirow{3}{*}{Median}}
			& Human & 0.065 & 0.127 & 0.091 & 0.156 \\
		\cline{2-6}
		\multicolumn{1}{ |c|  }{}
			& Sculpt & 0.091 & 0.169 & 0.127 & 0.183 \\ 
		\cline{2-6}
		\multicolumn{1}{ |c|  }{}
			& \textbf{\% Increase} & \textbf{40.2} & \textbf{33.5} & \textbf{40.4} & \textbf{17.4} \\ \hline
	\end{tabular}
\end{center}
\caption{\label{tab:humvsculpt}
Measures of the depth of the eye sockets on scans of humans vs. sculptures.}
\end{table}
		
For each aggregate measure, the sculpted scans had significantly deeper eye sockets than the human scans. Note that all fiducial points were identified by hand, the quality and resolution of the scans varies significantly, and the measures we propose are somewhat ad hoc. Furthermore, the size of the sample is much too small to draw concrete conclusions. Nonetheless we see a strong suggestion that sculptors such as Rodin routinely exaggerated features of facial anatomy such as the depth of eye sockets.

\section{Approach}
\label{sec:approach}
Our goal is to allow a user to interactively stylize an input face model guided by sculptural abstraction principles. Our system first computes an \emph{abstracted mesh} from the input face model. The abstracted mesh needs to convey both high-level abstract elements of the input mesh and semantically meaningful facial features. To this end, we use a \emph{Planes of the Head} model as our template to guide our abstracted mesh generation. 
The abstracted mesh serves as the \emph{sculptural abstraction framework} for the subsequent stylization steps, and is divided into meaningful regions corresponding to the planes from the template model.

The subsequent sculptural stylization optimization and stylization transfer phases of our system run at real-time rates and are designed for interactive use by a user in control of global and/or local parameter settings.
We stylize the abstracted mesh by minimizing an energy function with several terms of various purposes: to exaggerate angles between different regions; to enforce flatness within each segment; to regularize the stylization towards the original shape; and to enforce \emph{Lanteri constraints} that preserve geometric measurements characteristic of a person's identity.
The transfer of these stylizations back to the input mesh
produces real-time full resolution results for inspection and further adjustment. The amount of smoothing can be controlled here both globally and locally; and the interactive display of the full resolution deformed mesh is crucial in permitting fine user control over subtle deformations.

\subsection{Abstracted Mesh Generation}

We first generate the abstracted mesh $\mathcal{M_C}$ that captures locally meaningful sculptural abstractions from the input mesh $\mathcal{M}$. For general 3D models, we find that Variational Shape Approximation (VSA) \cite{Cohen-Steiner:2004} often derives plausible geometric abstractions from the input mesh by grouping faces of similar normals into contiguous regions. However for human faces, due to our heightened perceptions, VSA can cause noticeable artifacts because it does not preserve semantically meaningful segmentations and salient features such as eyes and lips. 

\begin{figure}
\centering

\subfloat{\includegraphics[width=0.3\linewidth]{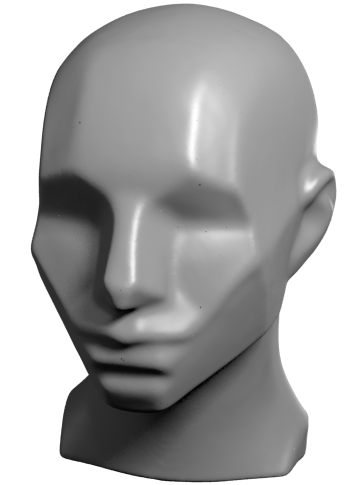}}\hspace{10px}
\subfloat{\includegraphics[width=0.3\linewidth]{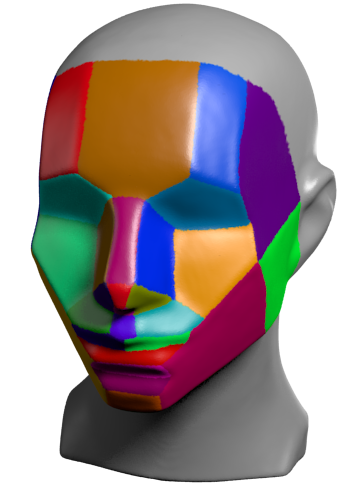}}
\caption{\label{fig:planesofthehead}\emph{Planes of the Head: Memorized} scan and segmentation.}

\subfloat[]{\includegraphics[width=0.32\linewidth]{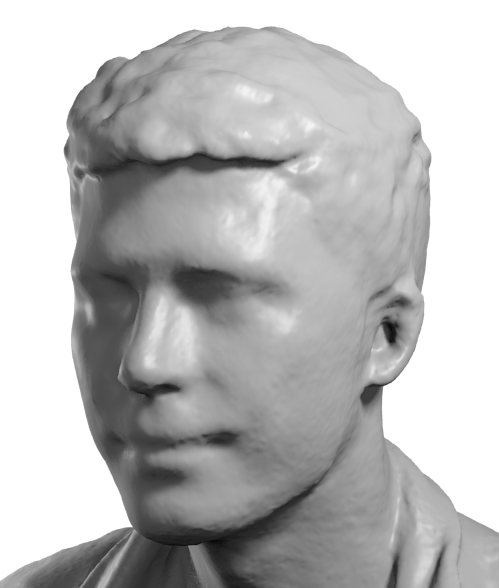}\label{fig:seg_original}}
\subfloat[]{\includegraphics[width=0.32\linewidth]{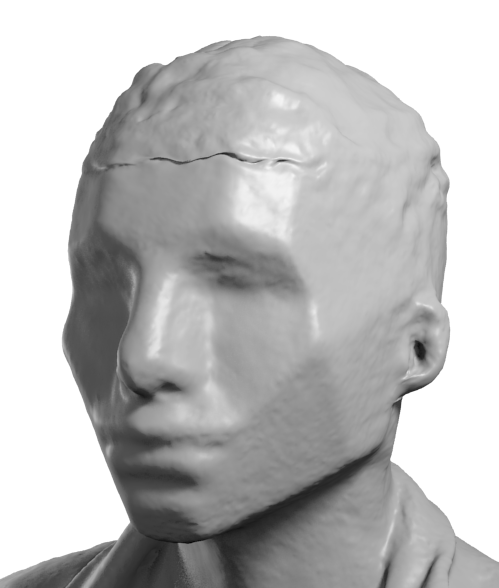}\label{fig:seg_aligned}}
\subfloat[]{\includegraphics[width=0.32\linewidth]{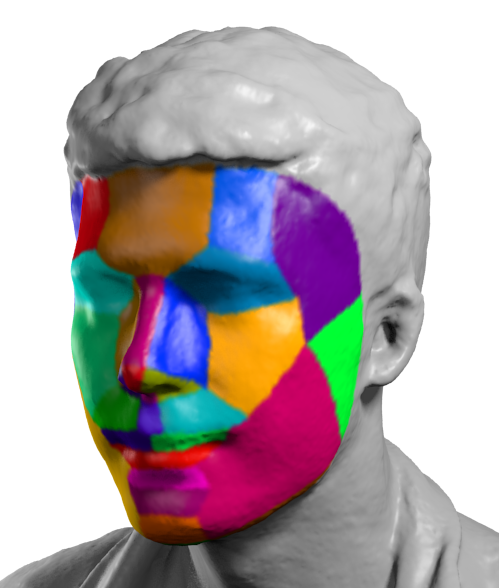}\label{fig:seg_segmented}}
\caption{\label{fig:segment}The segmentation of the template model is transferred to the input mesh using nonrigid alignment.}
\end{figure}

Thus, we employ a semantically segmented human head template to generate the abstracted meshes from human face models. More specifically, we use the \emph{Planes of the Head} model (Figure~\ref{fig:asaro}) because it takes into account both the principles of sculptural abstraction and facial semantics. To acquire the 3D template, we scanned a Planes of the Head model using KinectFusion~\cite{Newcombe:etal:ISMAR2011} and manually segmented the model into sculptor's planes (shown with colored segments in Figure~\ref{fig:planesofthehead}). For each novel input mesh $\mathcal{M}$ we transfer the segmentation by non-rigidly aligning $\mathcal{M}$ to the template, using the method of Li~\etal~\shortcite{li09robust}, and taking the nearest vertices as correspondences (Figure~\ref{fig:seg_aligned}).

The transferred segmentation divides $\mathcal{M}$ into $K$ \emph{regions} $\lbrace R_r \rbrace^K_{r=1}$ ($K=32$). We generate the abstracted mesh $\mathcal{M_C}$ by approximating each sculptor's plane with a small number of triangles. We follow the approach of Cohen-Steiner \etal~\shortcite{Cohen-Steiner:2004}: anchor vertices are placed along the borders between sculptor's planes and a constrained Delaunay triangulation is used to fill the regions. Note that our algorithm only uses the anchor vertices on the boundaries between regions, the triangulation is used for visualization purposes.

\subsection{Abstracted Mesh Stylization}
\label{sec:coarse}
\label{sec:plane_proxies}

After we compute the abstracted mesh $\mathcal{M_C}$ for the input mesh $\mathcal{M}$, we build up a stylization framework on the abstracted mesh in order to eventually transfer the stylization to the input mesh (Sec.~\ref{sec:transfer}). The stylization framework includes both \emph{exaggeration} of the angles between the regions and \emph{planarization} of the features within each region. Here we introduce a few definitions to facilitate our formulation of the stylization framework.

We first define the sculptor's plane $\pi_r$ approximating each region $R_r$ by the normal $\vect{n}_r$ and centroid $\vect{c}_r$. $\vect{c}_r$ is computed as the weighted average of the centroids of all the triangles within $R_r$ and $\vect{n}_r$ as the average of the triangle normals weighted by the triangle areas within $R_r$:
\begin{equation}
\vect{n}_r = \frac{\sum_{t\in R_r} A_t \vect{n}_t } {\left| \sum_{t\in R_r} A_t \vect{n}_t \right|}
~.
\label{eq:normal}
\end{equation} 
Since $A_t\vect{n}_t=\vect{v}_{t1} \times \vect{v}_{t2} + \vect{v}_{t2} \times \vect{v}_{t3} + \vect{v}_{t3} \times \vect{v}_{t1}$ where $\vect{v}_{t1}, \vect{v}_{t2}, \vect{v}_{t3}$ are the vertices of triangle $t$,  all the terms with respect to the internal edges in Eqn.~\ref{eq:normal} cancel out given that $\vect{v}_i \times \vect{v}_j = - \vect{v}_j \times \vect{v}_i$. Thus we only need to include the terms with respect to each boundary edge $e=(\vect{v}_{e1},\vect{v}_{e2})\in \partial R_i$:
\begin{equation}
\vect{n_r} = \frac{\sum_{e \in \partial R_r} \vect{v}_{e1} \times \vect{v}_{e2} }{\left|\sum_{e \in \partial R_r} \vect{v}_{e1} \times \vect{v}_{e2} \right|}
~.
\label{eq:normal_final}
\end{equation}
To facilitate the stylization transfer in (Sec.~\ref{sec:transfer}), we define an affine transformation $\vect{T}_r$ for each $R_r$ as follows:
\begin{equation}
\vect{T}_r = \vect{T}^r_r  \vect{T}^p_r,
\label{eq:affine}
\end{equation}
where $\vect{T}^r_r$ defines the rigid transformation implied by the change of $\vect{n}_r$ to $\vect{n}_r'$ and $\vect{c}_r$ to $\vect{c}_r'$ in the stylization optimization in this section:
\begin{equation}
\vect{T}^r_r = \left[ \vect{R}(\vect{n}_r, \vect{n}_r') \,\vert\, \vect{c}_r' - \vect{R}(\vect{n}_r, \vect{n}_r') \vect{c}_r \right]
\label{eq:affine_rigid}
~,
\end{equation}
where $\vect{R}(\vect{n}_r, \vect{n}_r')$ is the rotation matrix to transform $\vect{n}_r$ to $\vect{n}_r'$. This can be derived from Rodrigues' formula to compute the rotation matrix based on the axis $\vect{n}_r \times \vect{n}_r'$ and angle $\cos^{-1}(\vect{n}_r \cdot \vect{n}_r')$.

The planarization transformation $\vect{T}^p_r$ scales the affine space in the $\vect{n}_r$ direction relative to $\vect{c}_r$:
\begin{equation}
\vect{T}^p_r(\mu) = \left[ \vect{I}_3 - \mu \vect{n}_r\vect{n}_r^\top \,\vert\, \mu(\vect{n}_r\cdot \vect{c}_r) \vect{n}_r \right]
~,
\end{equation}
where $\mu$ is a user-defined variable which determines the amount of planarization and $\vect{I}_3$ is the 3x3 identity matrix. If a point $\vect{p}$ is at a distance $d$ from the plane then $\vect{T}^{p}_r(\mu)\vect{p}$ will be a distance $(1-\mu)d$ from the plane $\pi_r$.

Note that we can derive all the quantities mentioned above from vertex positions. Our optimization in the following will be formulated in terms of vertex positions directly or indirectly.

\subsubsection*{Energy formulation} We pose the stylization of our abstracted mesh as an energy minimization problem. We consider two types of energies in our formulation. The stylization energy exaggerates the angles between adjacent regions while keeping the regions themselves roughly planar; and the regularization energy keeps the solution close to the original abstracted mesh and the Lanteri constraints as suggested by Lanteri \shortcite{Lanteri:1985}. The total energy to be minimized is the sum of all stylization and regularization energies:
\begin{equation}
E = E_{style} + E_{reg}
\label{eq:opt}
~.
\end{equation}
\subsubsection*{Stylization energy} The stylization energy consists of two terms for exaggeration between adjacent regions $R_i$ and $R_j$ and planarization within each region:
\begin{equation}
E_{style} = \lambda_d\sum_{R_i\sim R_j} w_{i,j} \vect{n}_i\cdot \vect{n}_j
+ \lambda_f \sum_r \sum_{\vect{v}\in R_r} \left( \vect{n}_r\cdot \left( \vect{c}_r - \vect{v} \right) \right)^2.
\label{eq:style}
\end{equation}
Here, $R_i\sim R_j$ means that region $R_i$ shares a boundary with region $R_j$.
Weights $w_{i,j}$  scale the amount of exaggeration between $R_i$ and $R_j$. $\lambda_d$ ($d$ for dihedral) controls the overall amount of exaggeration while $\lambda_f$ controls how flat the regions should be.

The weights $w_{i,j}$ are set as the boundary length between $R_i$ and $R_j$ normalized by the average boundary length. We find that this weighting scheme ensures
 that each edge's deformation is weighted approximately according to its visual impact on the final model. Our system allows the user to alter these weights by specifying a scale factor $s_{i,j}$ such that: $w_{i,j}\gets s_{i,j}w_{i,j}$.

We can transform the dot product between unit normal vectors $\vect{n}_i\cdot \vect{n}_j$ into an energy with a sum of squares form to facilitate the use of efficient solvers such as Levenberg-Marquardt:
\begin{equation}\textstyle
\vect{n}_i\cdot \vect{n}_j = \frac{1}{2}\lVert \vect{n}_i + \vect{n}_j \rVert ^2 - 1
~.
\end{equation}

\subsubsection*{Regularization energy} The regularization energy $E_{reg}$ includes the following terms: 
\begin{equation}
E_{reg} = E_{area} + E_{edge} + E_{vertex} + E_{normal} + E_{Lanteri}.
\end{equation}
The edge and area terms are as follows:
\begin{gather}
E_{area} = \lambda_{a}\sum_{i=1}^K \left( 1 - \frac{ A(R_r)}{ A^0(R_r)} \right)^2 \\
E_{edge} = \lambda_{e}\sum_{e\in \mathcal{M}_C} \left(1 -\frac{ \lvert e \rvert}{ \lvert e^0 \rvert} \right)^2
\label{eq:area_edge_term}
\end{gather}
where $A(R_r)$ is the area of $R_r$. $\lvert e \rvert$ is the length of edge $e$ and the sum only includes edges on the borders between regions. The superscript ``0'' denotes the initial value. By using ratios with the initial value we make the terms scale invariant. 

The vertex and normal terms are simple $L_2$ error metrics:
\begin{gather}
E_{vertex} = \frac{\lambda_{v}}{2\lvert \bar{e} \rvert^2} \sum_{v\in \mathcal{V_C}} \lVert v - v_0 \rVert^2 \\
E_{normal} = \frac{\lambda_{n}}{2}\sum_{i=1}^K \left\lVert \vect{n}_r - \vect{n}_r^0 \right\rVert ^2
\label{eq:vertex_normal_term}
\end{gather}
where $\lvert \bar{e} \rvert$ is the mean edge length of the mesh used to make the vertex term scale invariant. 

All of these regularization terms are needed to avoid various degeneracies of the energy function and maintain closeness to the original mesh, but the results are not critically affected by their weights. We used constant values of $\lambda_v$, $\lambda_n$, $\lambda_e$, and $\lambda_a$ (given in Section~\ref{sec:results}) for all results shown in this paper.

\subsubsection*{Lanteri constraints}
\label{sec:Lanteri}

\begin{figure}
\centering
\def\imw{0.3\linewidth}
\hfill
\subfloat{\includegraphics[width=\imw]{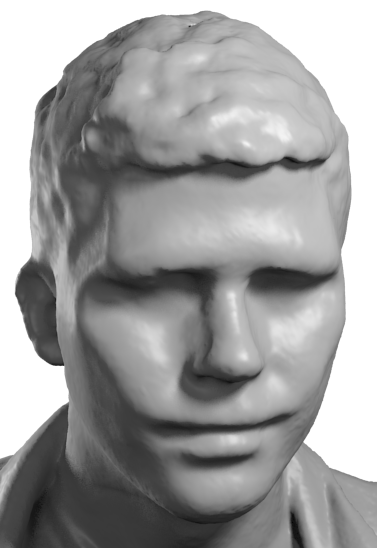}}
\hfill
\subfloat{\includegraphics[width=\imw]{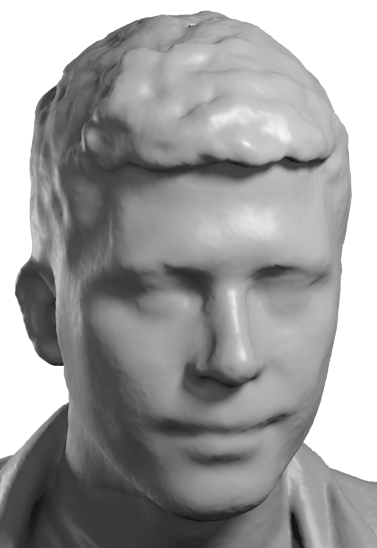}}
\hfill
\subfloat{\includegraphics[width=\imw]{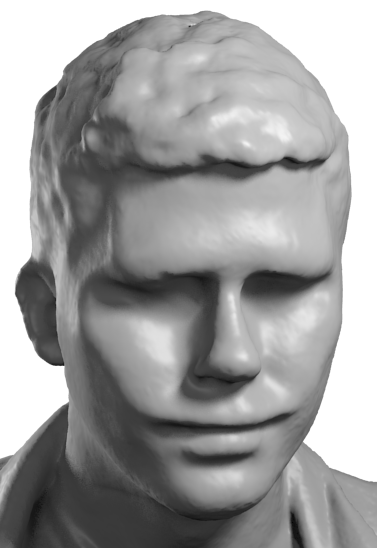}}
\hfill
\caption{Lanteri constraints help to retain the personality of the original mesh (middle) in the stylized mesh (left). Without the constraints unwanted deformations can occur such as the eyes getting closer together (right).}
\label{fig:lanteri}
\end{figure}

Based on the writings of Edouard Lanteri, we posit that certain relative measurements on the face are critical to maintain the individual ``personality'' of the model (Figure \ref{fig:lanteri}). We thus consider three types of \emph{Lanteri constraints} to formulate the Lanteri energy $E_{Lanteri}$. An absolute position constraint (Eq.~\ref{eq:lanteri:abs_pos}) ensures that certain points of interest remain at their original positions, a relative position constraint (Eq.~\ref{eq:lanteri:rel_pos}) ensures that the relative position of two points remains the same, and a relative distance constraint (Eq.~\ref{eq:lanteri:rel_dis}) ensures a constant distance between two points during stylization. The energy is formulated as a sum of term of the following form:
%
\begin{subequations}
\begin{gather}
\sum_{\vect{p}} \frac{\lambda_{v}}{2\lvert \bar{e} \rvert^2}
\lVert \vect{T}(\vect{p})\vect{p} - \vect{p}\rVert^2 \label{eq:lanteri:abs_pos}\\
\sum_{\vect{p_1,p_2}} \frac{\lambda_{v}}{2}
\frac{
\lVert \left(\vect{T}(\vect{p}_1)\vect{p}_1 - \vect{T}(\vect{p}_2)\vect{p}_2 \right) - \left( \vect{p}_1 - \vect{p}_2 \right) \rVert^2}
{\lVert \vect{p}_1-\vect{p}_2 \rVert^2} \label{eq:lanteri:rel_pos}\\
\sum_{\vect{p_1,p_2}} \frac{\lambda_{v}}{2}
\frac{
\left( \lVert\vect{T}(\vect{p}_1)\vect{p}_1 - \vect{T}(\vect{p}_2)\vect{p}_2 \rVert - \lVert \vect{p}_1 - \vect{p}_2  \rVert \right)^2}
{\lVert \vect{p}_1-\vect{p}_2 \rVert^2}\label{eq:lanteri:rel_dis}
\end{gather}
\end{subequations}
%
$\vect{T}(\vect{p})$ transforms $\vect{p}$ based on its closest regions' affine transformation (Sec.~\ref{sec:transfer}). The specific feature points used to formulate these constraints are given in Appendix \ref{app:Lanteri}. 

\subsubsection*{Energy minimization} To find the minimum of Equation~\ref{eq:opt} the energy is written as a function of the vertex positions and minimized using the Levenberg-Marquardt method in Ceres Solver~\cite{ceres-solver}. The vertices on the open boundary of the mesh are kept constant during the optimization. Standard thresholds on function tolerance and gradient tolerance are used to determine convergence.

\subsection{Stylization Transfer}
\label{sec:transfer}

To transfer stylization of the abstracted mesh back onto the input mesh, we need to ensure smooth transitions between different regions to avoid any visible artefacts. We define a set of skinning weights $w_{i,r}$ for each vertex $\vect{v}_i$ with respect to each region $R_r$. The final affine transformation $\vect{T}(\vect{v}_i)$ for each vertex $\vect{v}_i$ of the input mesh is defined as the weighted average of all relevant $\vect{T}_r$:
\begin{equation}
\vect{T}(\vect{v}_i) = \sum_{r=0}^K w_{i,r} \vect{T}_r,
\end{equation}
where we define $\vect{T}_0$ to be the identity transformation which is assigned to the region outside of the optimization area to enable smoothing over this boundary.

\begin{figure}[t]
\centering
\def\imw{0.21\linewidth}\hfill
\subfloat{\includegraphics[width=\imw]{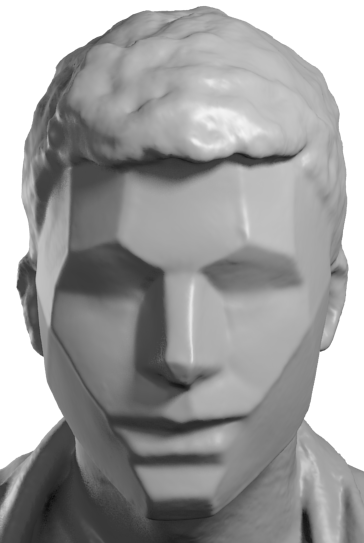}}\hfill
\subfloat{\includegraphics[width=\imw]{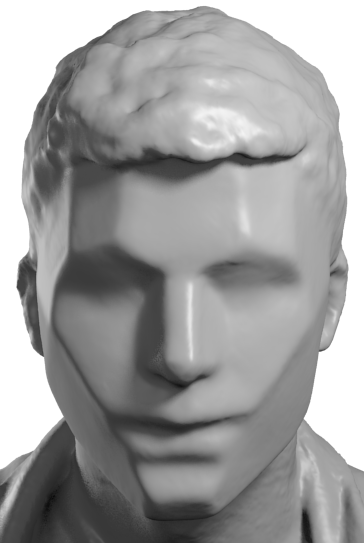}}\hfill
\subfloat{\includegraphics[width=\imw]{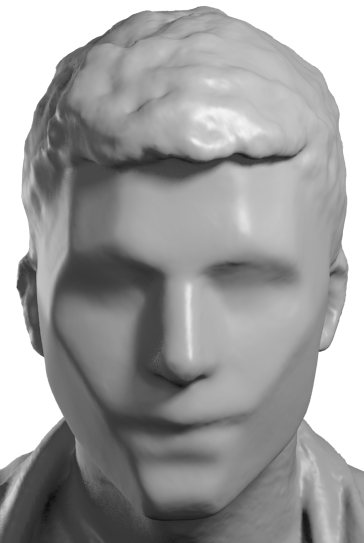}}\hfill
\subfloat{\includegraphics[width=\imw]{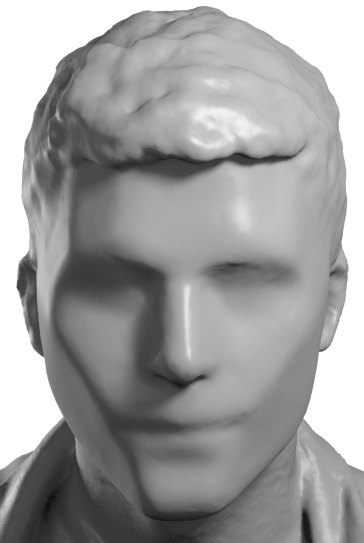}}\hfill
\caption{\label{fig:smooth}Varying levels of stylization smoothness can enhance or smooth the boundaries between sculptors' planes while the orientation of the planes remains the same.}
\end{figure}

Our system allows the user to adjust the stylization smoothness across different regions. Figure \ref{fig:smooth} shows one example where the user creates different stylization smoothness using our interactive user interface. To efficiently compute the skinning weights for interactive stylization smoothness editing, we pre-compute a pyramid of skinning weights $w_{i,r}^l$ with different levels of smoothing. Given a smoothing scale $s$, we can then linearly interpolate the proper $w_{i,r}$ between the two closest levels of skinning weights $w_{i,r}^l$ and $w_{i,r}^{l+1}$.

The first level of the pyramid contains the original, un-smoothed skinning weights. In this case we set the weights $w_{i,r}$ to be the number of faces in $R_r$ that contain vertex $\vect{v}_i$, normalized so that $\sum_r w_{i,r} = 1$. For all vertices not on a region border there will only be a single non-zero weight. To compute the skinning weights at the next level, we perform simple Laplacian smoothing based on the adjacent skinning weights:
\begin{equation}
w_{i,r} \gets \frac{1}{\lvert \vect{v}_i\sim \vect{v}_j \rvert}\sum_{\vect{v}_i\sim \vect{v}_j} w_{j,r},
\end{equation}
where $\vect{v}_i\sim \vect{v}_j$ means that $\vect{v}_i$ is adjacent to $\vect{v}_j$. The smoothing increases as a geometric sequence in the number of smoothing iterations as we go down the pyramid.

After the pyramid has been precomputed, different amounts of smoothing can be interactively set for different areas of the mesh. To accomplish this at interactive rates, we adapt a method for controlling spatially varying blur from Diffusion Curves~\cite{diffusion-curves}: The user selects a boundary $\{i,j\}$ between regions $R_i$ and $R_j$ and assigns a smoothing scale $s$. All vertices on the boundary $\{i,j\}$ are then assigned the value $s$. The system propagates the smoothing scale within both regions by solving a Laplace equation with Dirichlet boundary condition within each region:
\begin{equation}
 \nabla^2 s = 0
\end{equation}
where the Laplacian operator is discretized by the cotangent Laplacian weights. This system is solved using a Cholesky decomposition, and the solution is used to look up the smoothed weights from the skinning weight pyramid. Since the regions are typically small (two regions), and the Cholesky decomposition is independent of the value of $s$, the updates are very fast.

\subsection{Interaction}

\begin{figure}
\centering
\hfill
\subfloat{\includegraphics[width=0.49\linewidth]{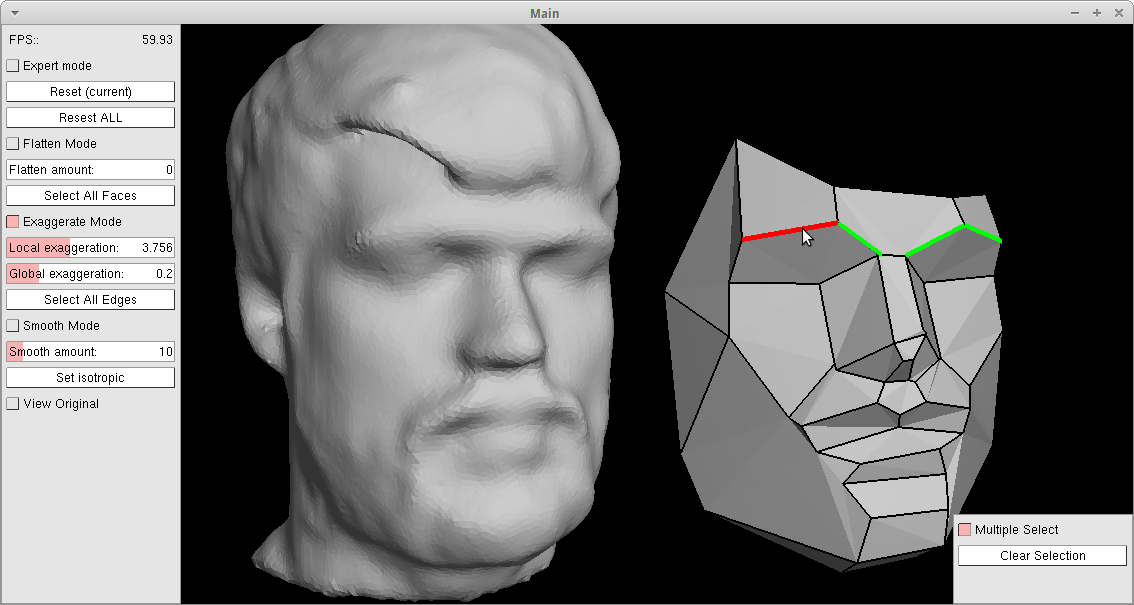}}
\hfill
\subfloat{\includegraphics[width=0.49\linewidth]{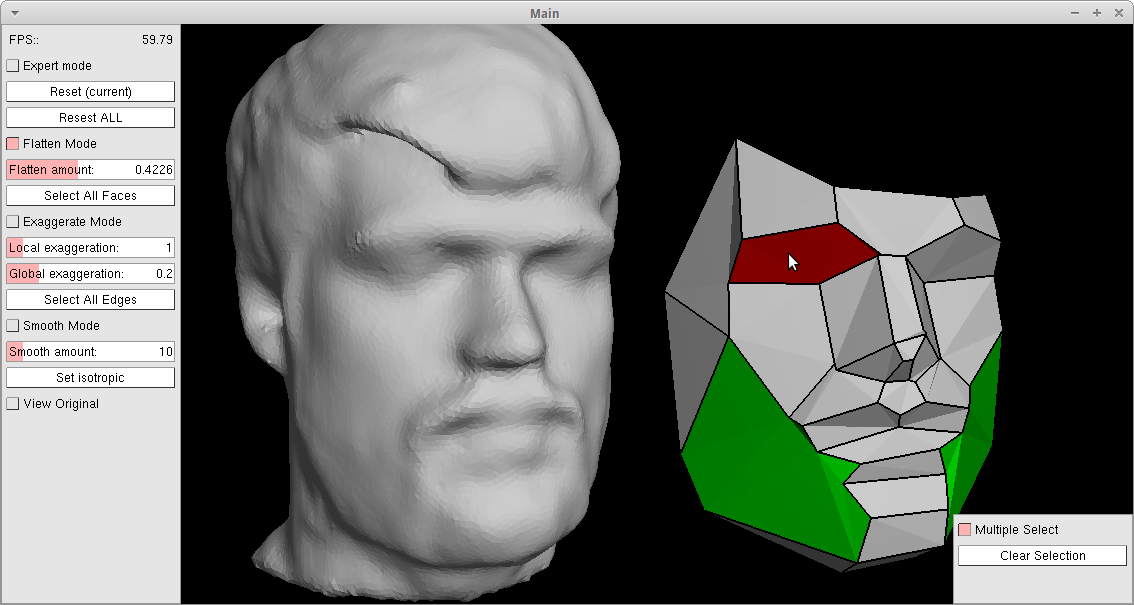}}
\hfill
\caption{A user can adjust intuitive exaggeration and smoothing terms for specific edges (left) or planarization terms for specific faces (right) of the abstracted mesh. The resulting deformations are shown interactively on the full resolution mesh.}
\label{fig:interaction}
\end{figure}

\begin{figure}
\centering
\def\imw{0.2\linewidth}
\subfloat{\includegraphics[width=\imw]{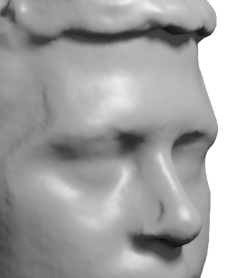}}
\subfloat{\includegraphics[width=\imw]{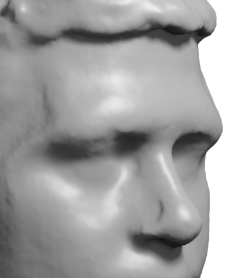}}
\subfloat{\includegraphics[width=\imw]{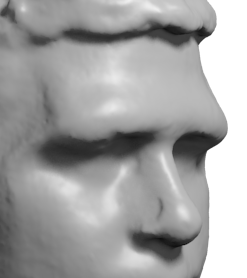}}
\subfloat{\includegraphics[width=\imw]{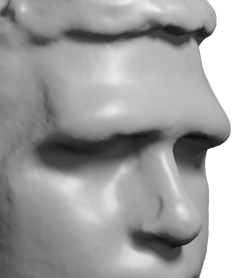}}
\subfloat{\includegraphics[width=\imw]{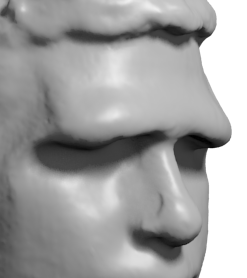}}
\caption{Our system allows for local as well as global controls. Here, we show increasing exaggeration of the brow.}
\label{fig:brow}
\end{figure}

Once we have aligned a new scan with the template model and produced its abstracted mesh, our full optimization framework for abstracted mesh stylization (Sec.~\ref{sec:plane_proxies}) and stylization transfer (Sec.~\ref{sec:transfer}) runs at an interactive frame-rate of 10Hz on a standard desktop workstation. Our GUI (Fig.~\ref{fig:interaction}) provides users with intuitive controls for the amount of exaggeration, the amount of planarization and transform smoothness. Each of these can be adjusted globally, or for more detailed control, exaggeration and smoothness amounts can be adjusted for each edge (Fig.~\ref{fig:brow}) and planarization amounts can be adjusted for each face. These correspond to intuitive visual changes, modifying the visual contrast between regions, displayed live in the form of the full resolution deformation result. Please see the supplementary video for a screen-captured demonstration.

\begin{figure}[t]
\centering
\def\imw{0.21\linewidth}
\captionsetup[subfigure]{labelformat=empty}\hfill
\subfloat[$\lambda_{d}=0.5$]{\includegraphics[width=\imw]{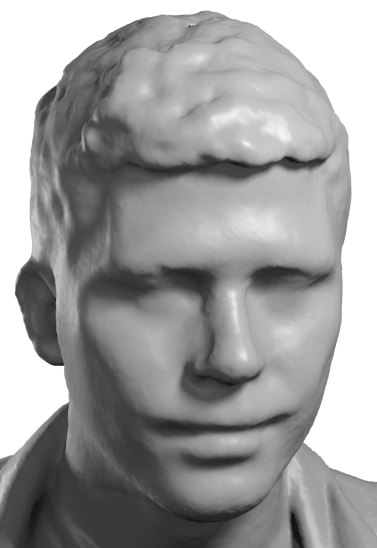}}\hfill
\subfloat[$\lambda_{d}=1.0$]{\includegraphics[width=\imw]{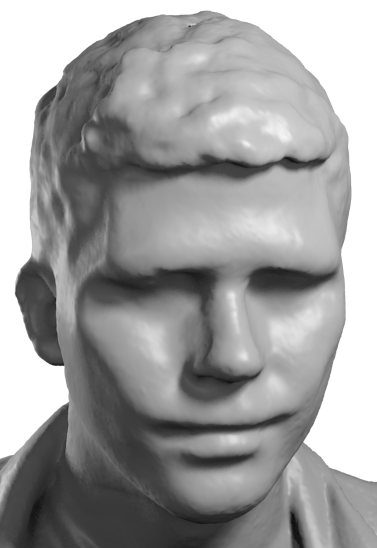}}\hfill
\subfloat[$\lambda_{d}=1.6$]{\includegraphics[width=\imw]{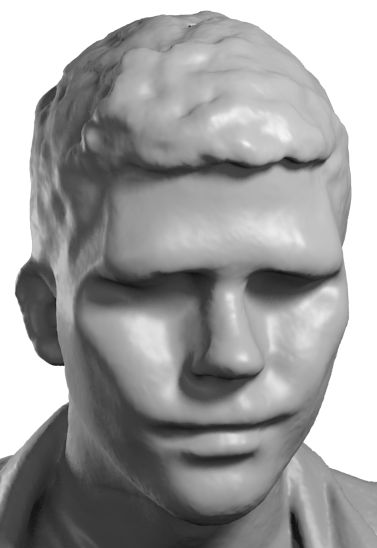}}\hfill
\subfloat[$\lambda_{d}=2.3$]{\includegraphics[width=\imw]{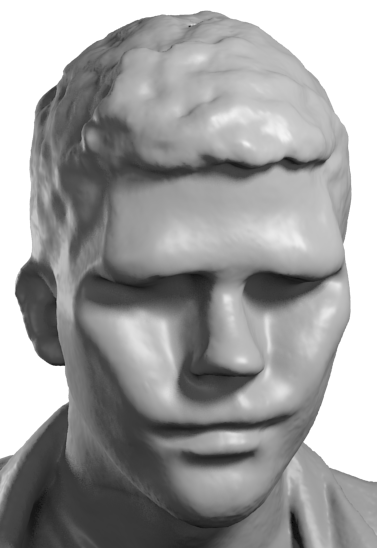}}\hfill
\caption{\label{fig:varying_abstraction}Increasing amounts of \textit{stylization}, from left to right. We intentionally push the exaggeration too far to demonstrate how the Lanteri constraints keep salient features in correct positions.}
\def\imw{0.15\linewidth}
\subfloat{\includegraphics[width=\imw]{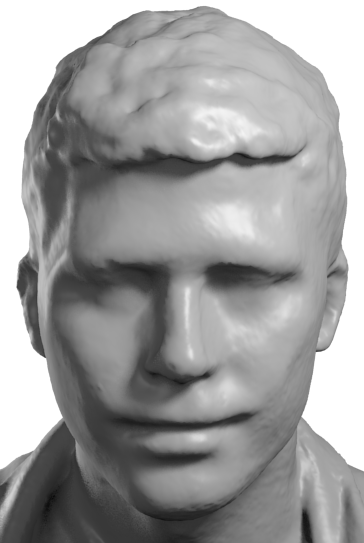}}
\subfloat{\includegraphics[width=\imw]{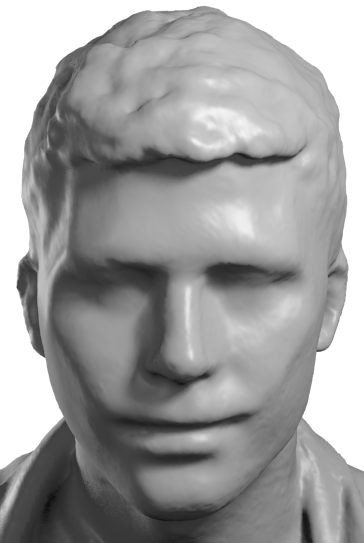}}
\subfloat{\includegraphics[width=\imw]{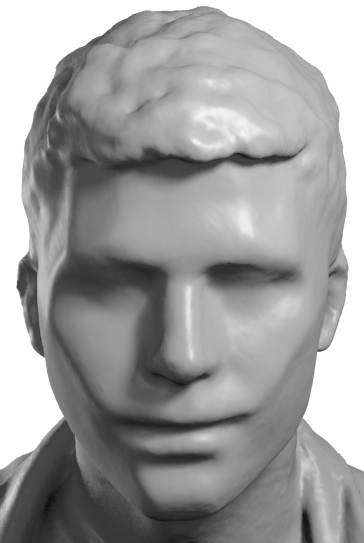}}
\subfloat{\includegraphics[width=\imw]{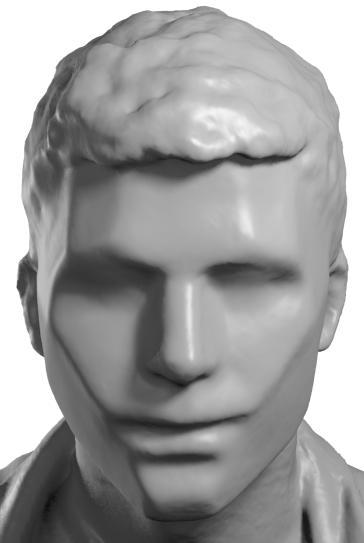}}
\subfloat{\includegraphics[width=\imw]{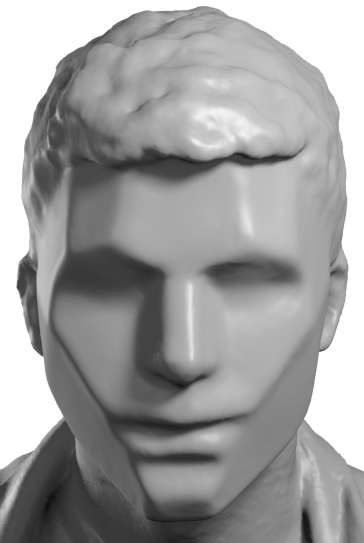}}
\subfloat{\includegraphics[width=\imw]{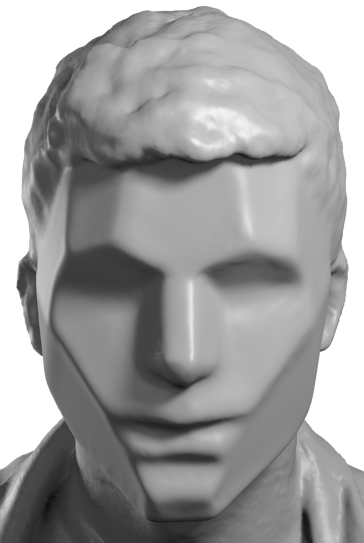}}
\caption{Increasing planarization (left to right)}
\end{figure}

\begin{figure}[t]
\centering
\hfill
\subfloat{\includegraphics[width=0.48\linewidth]{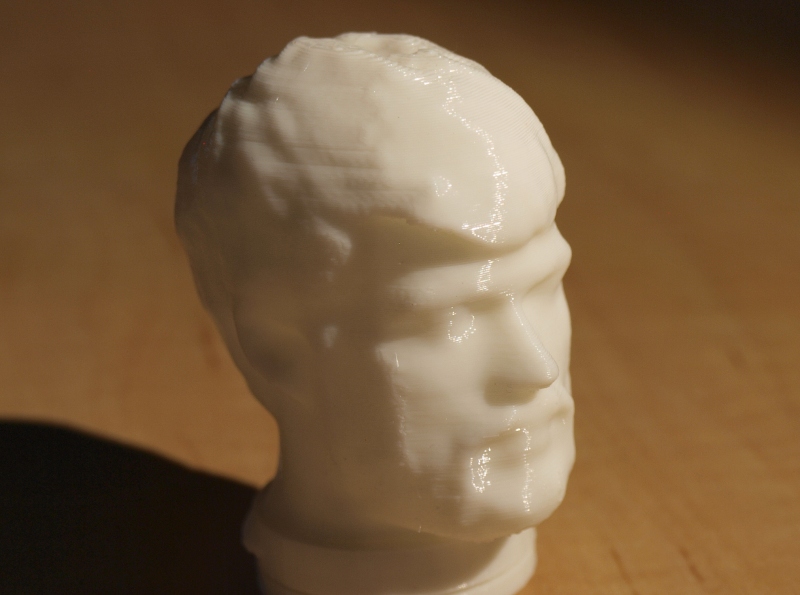}}
\hfill
\subfloat{\includegraphics[width=0.48\linewidth]{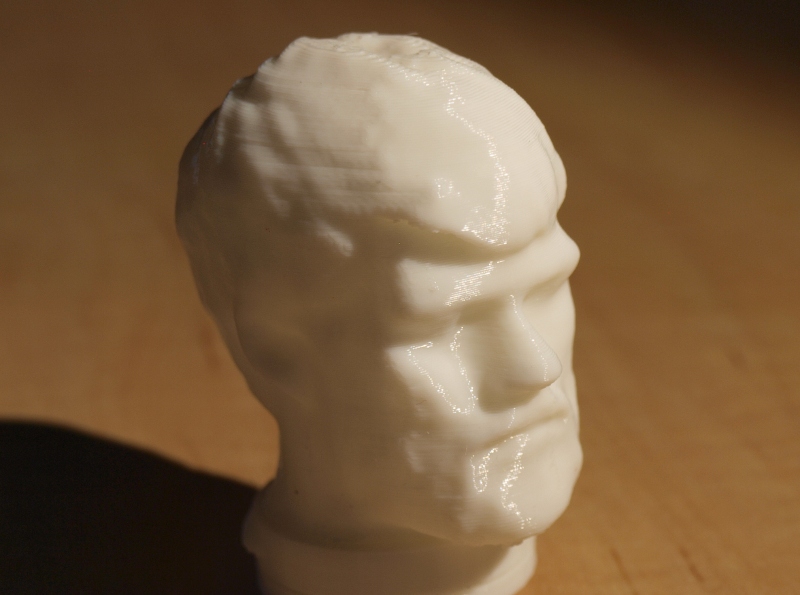}}
\hfill
\caption{3D printed busts --- original (left) vs. stylized (right).}
\label{fig:printed}
\end{figure}

\section{Results}
\label{sec:results}
The input triangular mesh can come from any source. The only requirement is that the areas to be abstracted are connected, manifold and contain no holes. We tested our system on meshes from two different sources; scans we made of a number of volunteers using the method of KinectFusion~\cite{Newcombe:etal:ISMAR2011}, and meshes obtained from a state-of-the-art multi-view stereo reconstruction system~\cite{Bee10}. 

\begin{figure}
\centering
\def\imw{0.3\linewidth} 
\hfill
\subfloat{	\includegraphics[width=\imw]{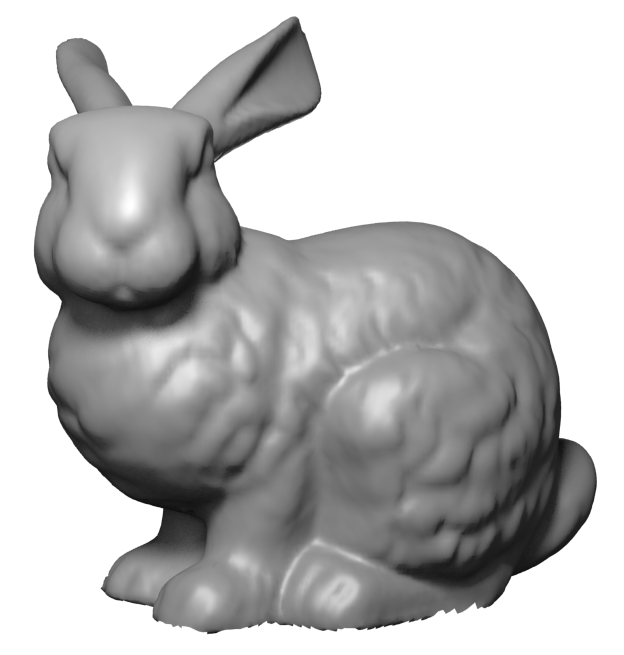}} \hfill
\subfloat{	\includegraphics[width=\imw]{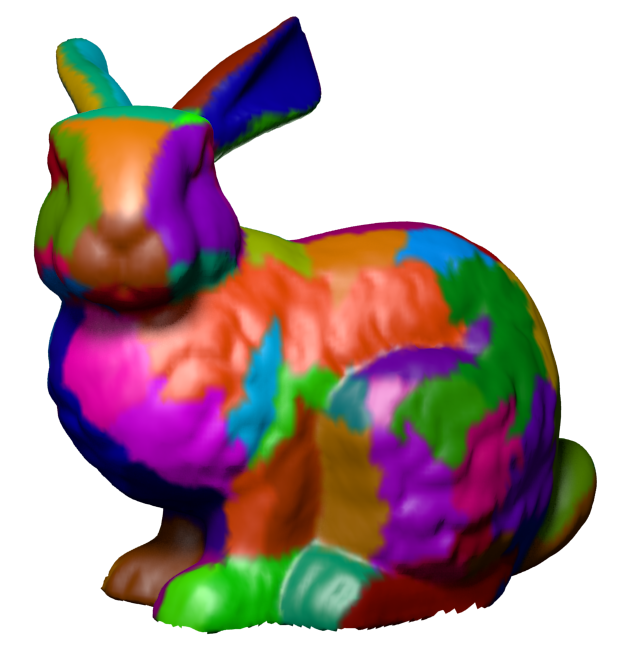}} \hfill
\subfloat{	\includegraphics[width=\imw]{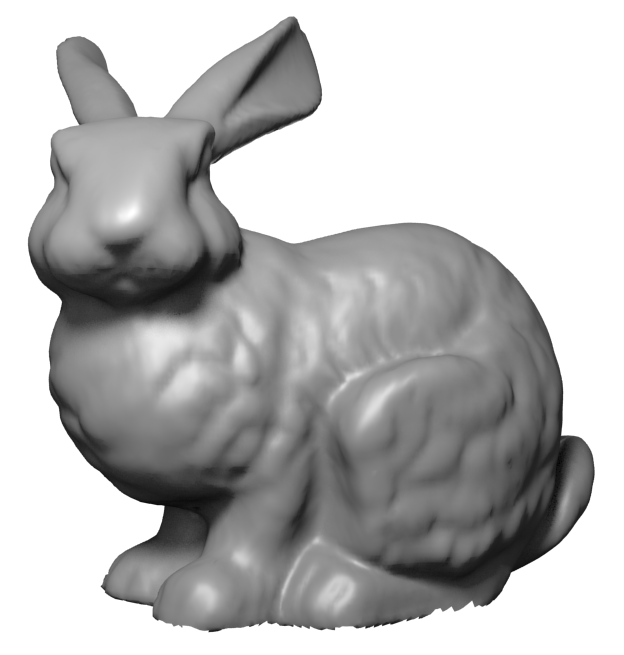}} \hfill \\
\hfill
\subfloat{	\includegraphics[width=\imw]{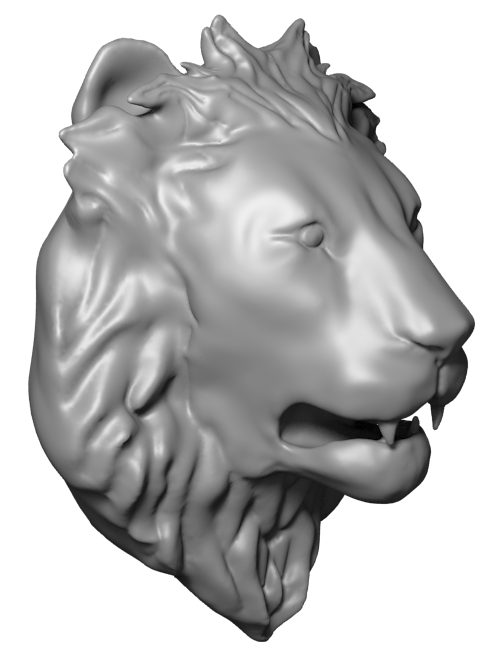}} \hfill
\subfloat{	\includegraphics[width=\imw]{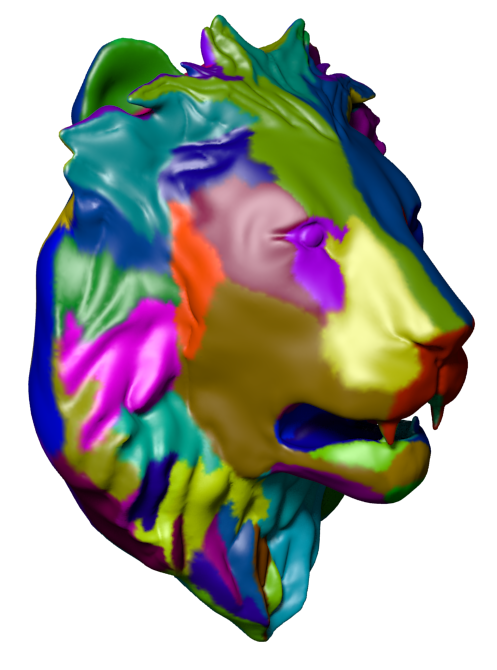}}\hfill
\subfloat{	\includegraphics[width=\imw]{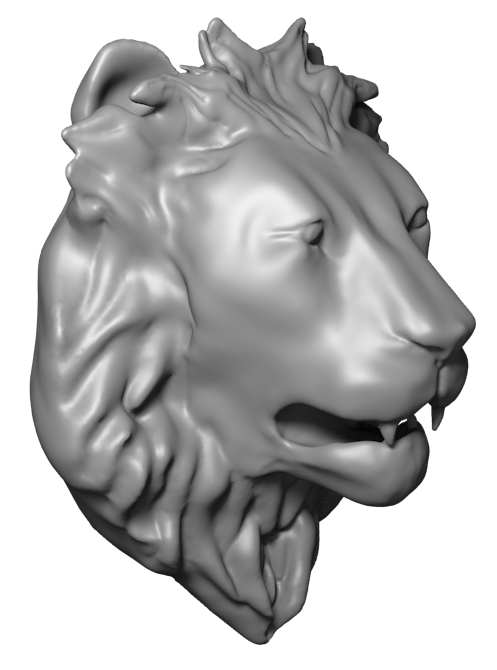}} \hfill
  \caption{Original mesh (left), segmentation (middle), and stylized mesh (right). Note the emphasis of facial features on the bunny\protect\footnotemark and sharper definition of the eye sockets on the lion\protect\footnotemark.}
\label{fig:bunny}
\end{figure}

The deformations produced by our method typically enhance cheekbones, deepen eye sockets, and emphasize the brows, chin
and lips (see Figure~\ref{fig:examples}). This approach is especially effective for the KinectFusion scans, which are sometimes lifeless and hard to recognize due to their lower fidelity. Further, materials used in 3D printing can be more translucent than skin, giving a blank look to printed busts. Our method gives such busts stronger definition by accentuating facial features, often making them more recognizable (see Figure~\ref{fig:printed}).
Our method can also be applied to models other than faces by using VSA for abstraction. In this case, the models are often thinned, and concavities become more pronounced (see
Figure~\ref{fig:bunny}). The deformations shown in these figures are subtle, by design, and may be hard to see clearly in the printed paper. We direct the reader to view the electronic version, and especially to our supplementary video, which demonstrates the abstractions more clearly by cutting rapidly between inputs and outputs.

For the stylization of the abstracted mesh (Section~\ref{sec:coarse}), the optimization is performed over a small number ($n<100$) of vertices on the coarse mesh, and thus takes only a fraction of a second, regardless of the input scan resolution. The stylization transfer step (Section~\ref{sec:transfer}) works with the full resolution mesh, but since only local linear solves and linear operations are required, both steps together can be run interactively. For an input mesh with 30k vertices our system runs at 10Hz on a Linux laptop, including all rendering, with room for further optimisation. For one of the Beeler~\etal meshes containing 375k vertices the system runs at about 2Hz. Currently the system runs entirely on the CPU, however, the stylization transfer step would be well suited to GPU acceleration.

Most of the weight parameters described in Sections~\ref{sec:coarse} and~\ref{sec:transfer} are held constant for all examples in this paper: $\lambda_{a} = 10$ and $\lambda_{e} = 4$, $\lambda_{v}=60$, $\lambda_{n}=1$, and $\lambda_{f}  = 1$. The default exaggeration weights yield pleasing results with the user just specifying $0\leq \lambda_{d}<3$ to control the overall amount of deformation. However, the user can customize the output by specifying per-edge exaggeration and smoothing weights and per-face planarization weights.

\addtocounter{footnote}{-1}
\footnotetext{Source: Stanford University Computer Graphics Laboratory}
\addtocounter{footnote}{1}
\footnotetext{Copyright \emph{purita}, licensed under CC BY-NC 3.0. Source: \url{http://www.thingiverse.com/thing:215420}}

\begin{figure}[t]
\centering
\def\imw{0.24\linewidth}
\vspace{-10px}
\subfloat{\includegraphics[width=\imw]{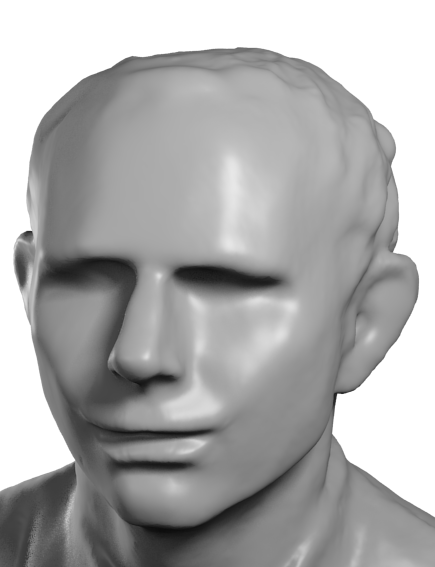}}
\subfloat{\includegraphics[width=\imw]{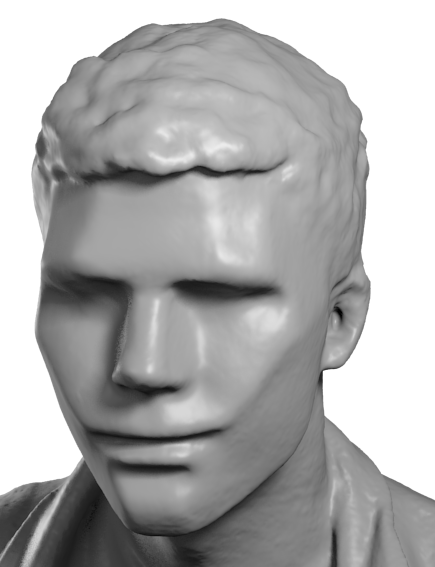}}
\subfloat{\includegraphics[width=\imw]{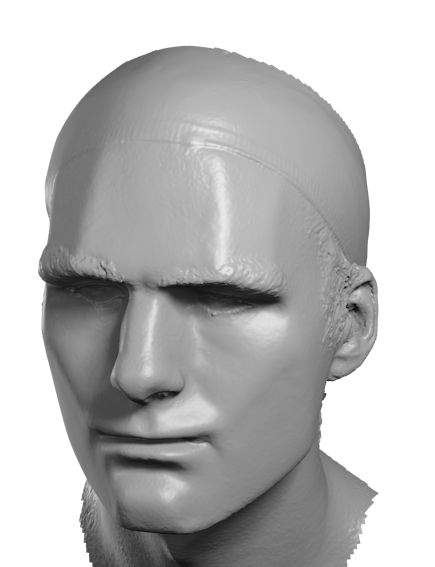}}
\subfloat{\includegraphics[width=\imw]{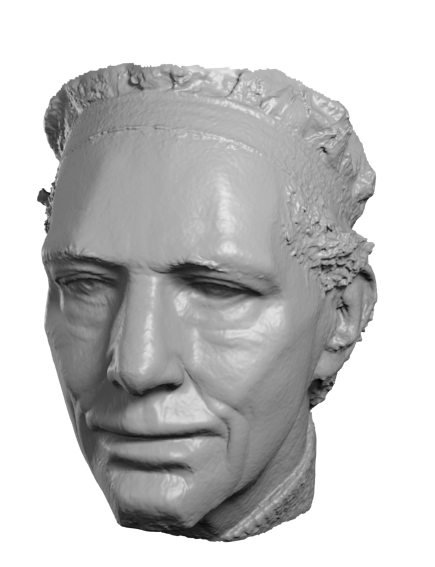}}\\
\vspace{-10px}
\subfloat{\includegraphics[width=\imw]{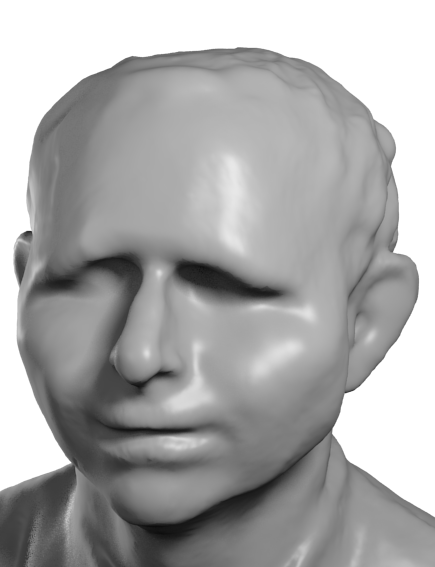}}
\subfloat{\includegraphics[width=\imw]{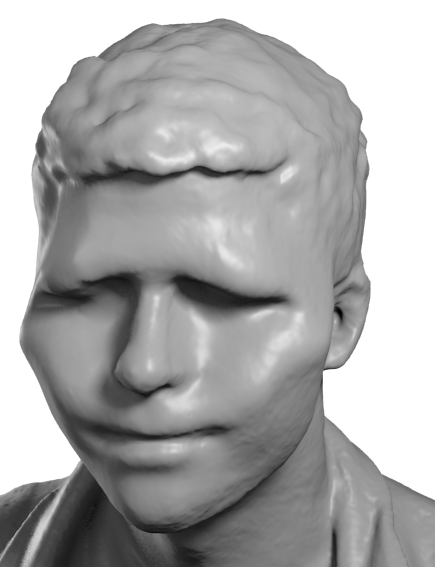}}
\subfloat{\includegraphics[width=\imw]{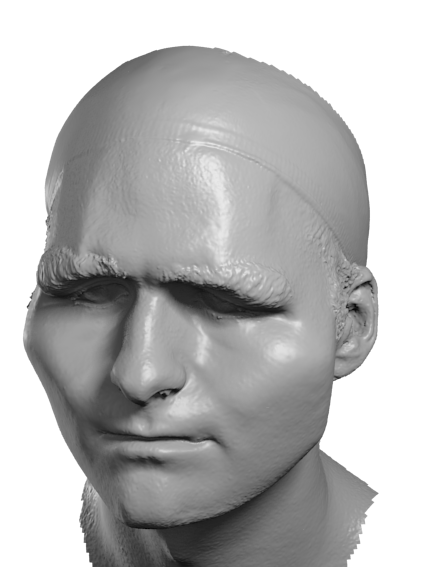}}
\subfloat{\includegraphics[width=\imw]{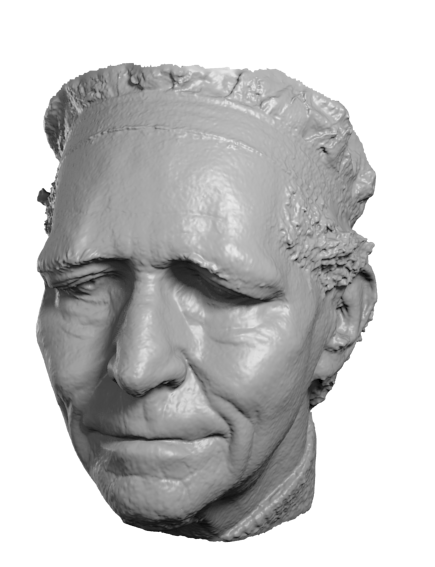}}\\
\caption{A style can be transferred to other meshes by using the same set of weights.}
\label{fig:same_settings}
\end{figure}

\begin{figure*}[ht]
\centering
\def\imw{0.165\linewidth}
\subfloat{\includegraphics[width=\imw]{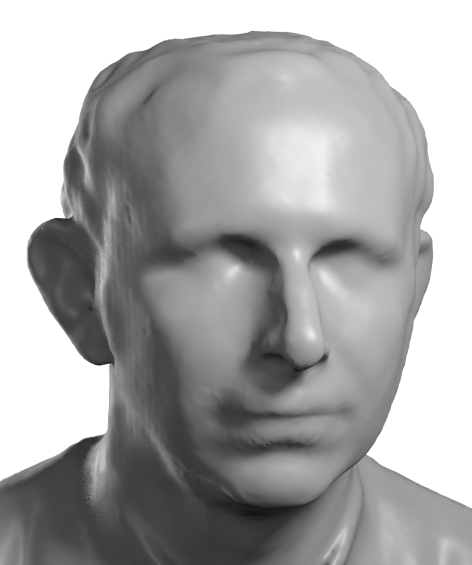}}
\hfill
\subfloat{\includegraphics[width=\imw]{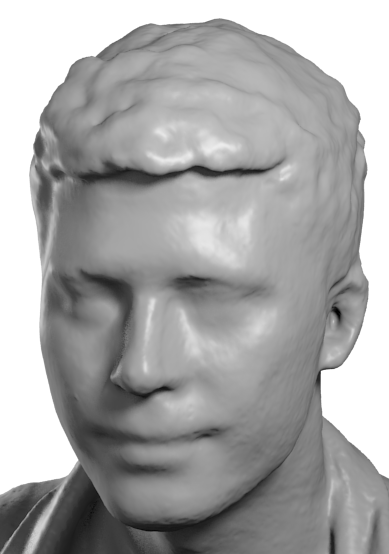}}
\hfill
\subfloat{\includegraphics[width=\imw]{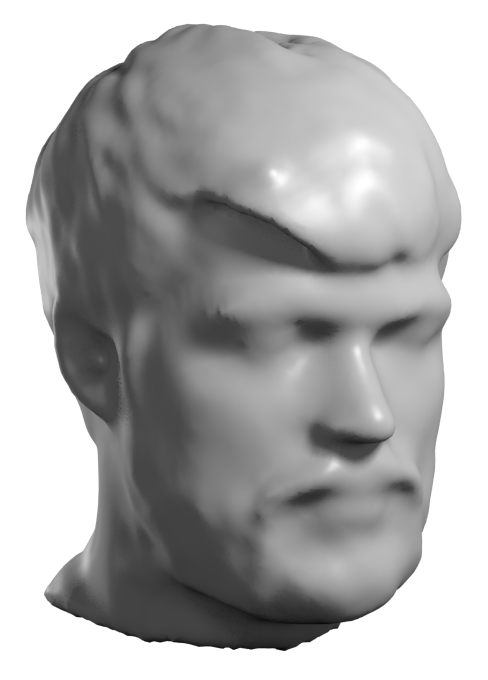}}
\hfill
\subfloat{\includegraphics[width=\imw]{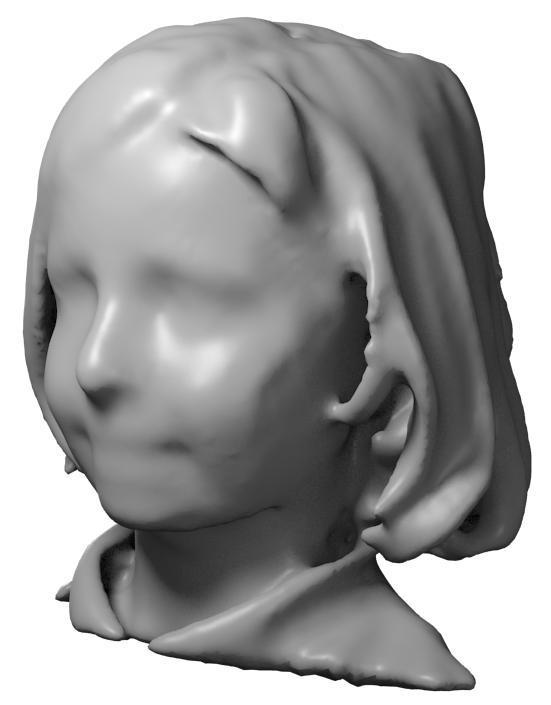}}
\hfill
\subfloat{\includegraphics[width=\imw]{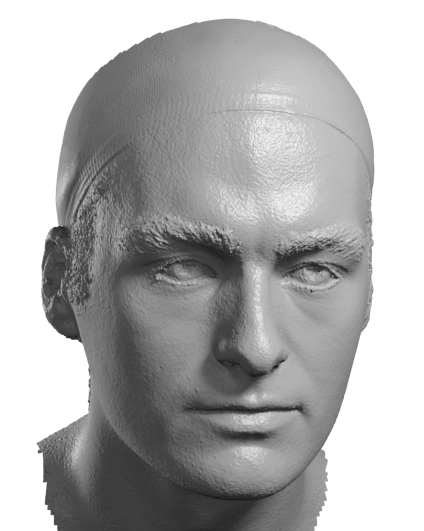}}
\hfill
\subfloat{\includegraphics[width=\imw]{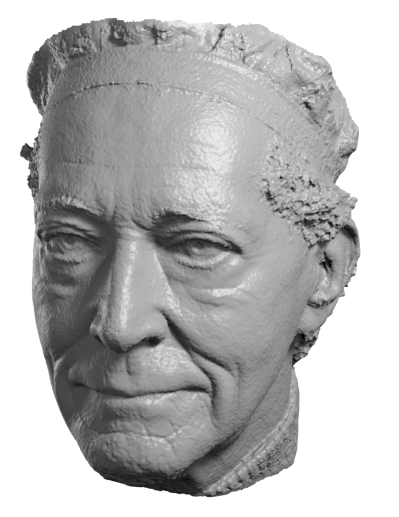}}\\
\vspace{-10px}
\setcounter{subfigure}{0}%
\subfloat[]{\includegraphics[width=\imw]{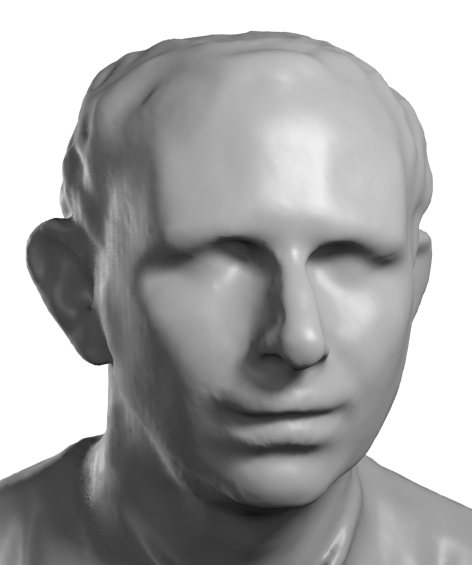}\label{fig:andy_ex}}
\hfill
\subfloat[]{\includegraphics[width=\imw]{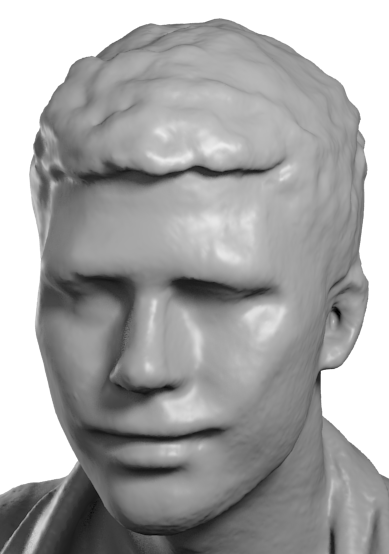}\label{fig:daniel_ex}}
\hfill
\subfloat[]{\includegraphics[width=\imw]{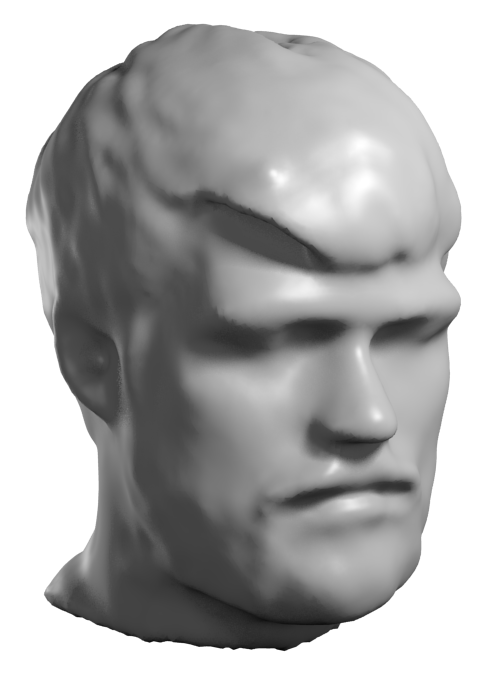}\label{fig:jan_ex}}
\hfill
\subfloat[]{\includegraphics[width=\imw]{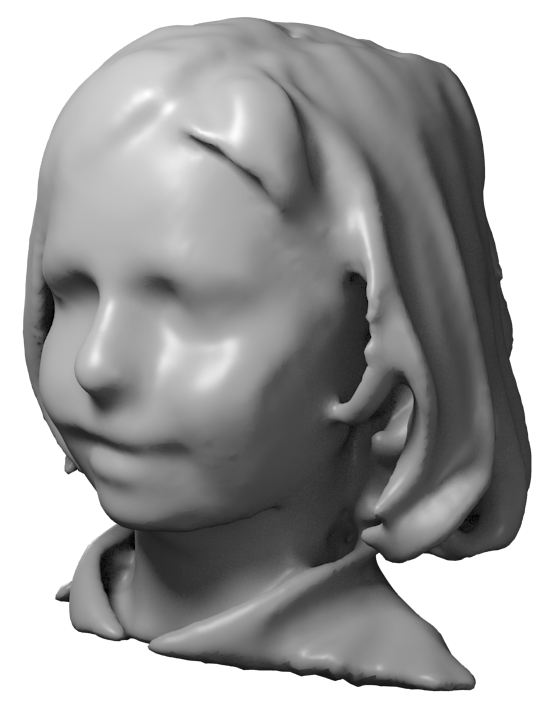}\label{fig:blanca_ex}}
\hfill
\subfloat[]{\includegraphics[width=\imw]{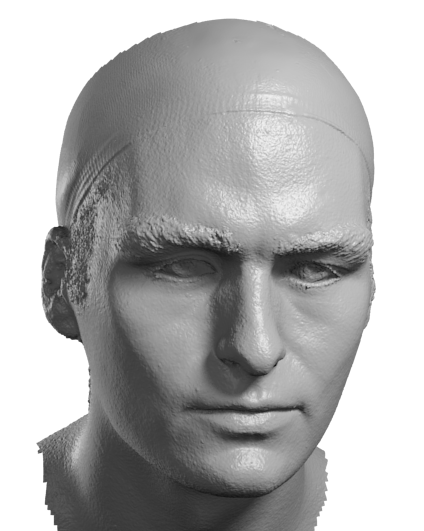}\label{fig:disney1_ex}}
\hfill
\subfloat[]{\includegraphics[width=\imw]{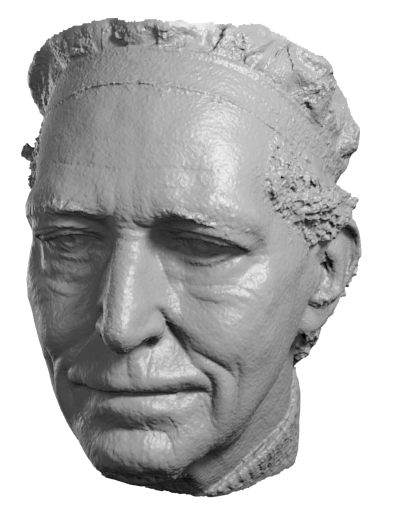}\label{fig:disney2_ex}}
\caption{(a)--(d) KinectFusion scans; (e),(f) scans from Beeler~\etal ~ Top row: input; bottom row: output.}
\label{fig:examples}
\def\imw{0.15\linewidth}
\subfloat[]{\includegraphics[width=\imw]{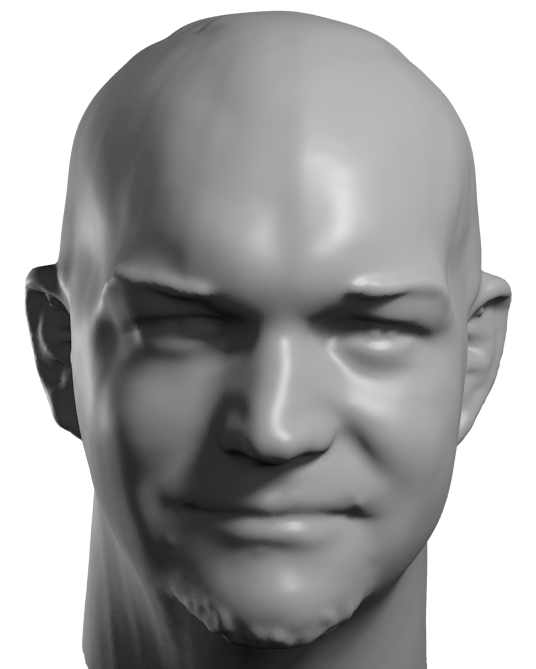}}
\subfloat[]{\includegraphics[width=\imw]{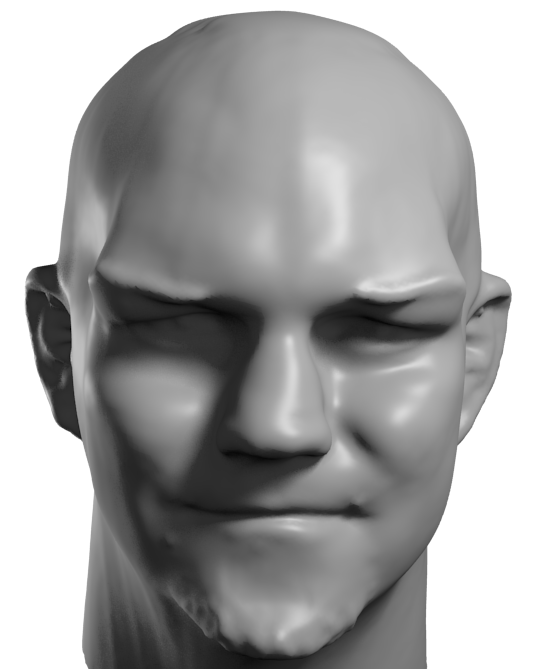}}
\subfloat[]{\includegraphics[width=\imw]{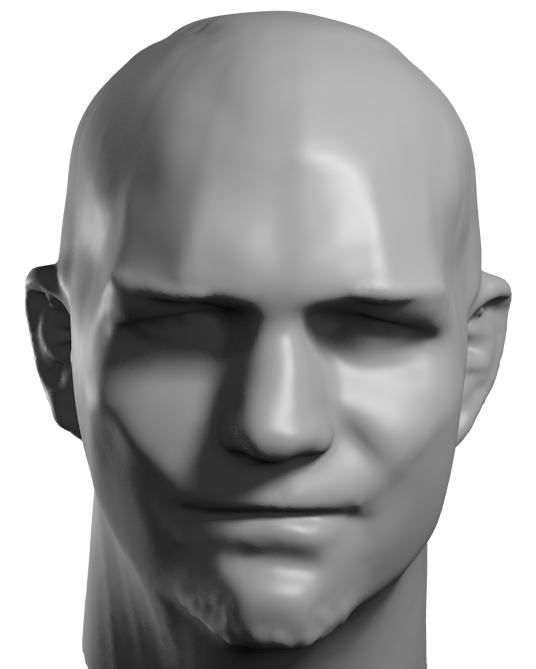}}
\subfloat[]{\includegraphics[width=\imw]{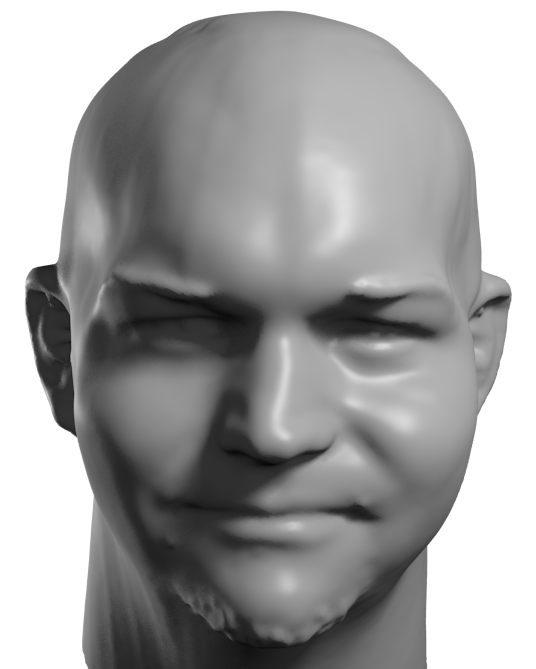}}
\subfloat[]{\includegraphics[width=\imw]{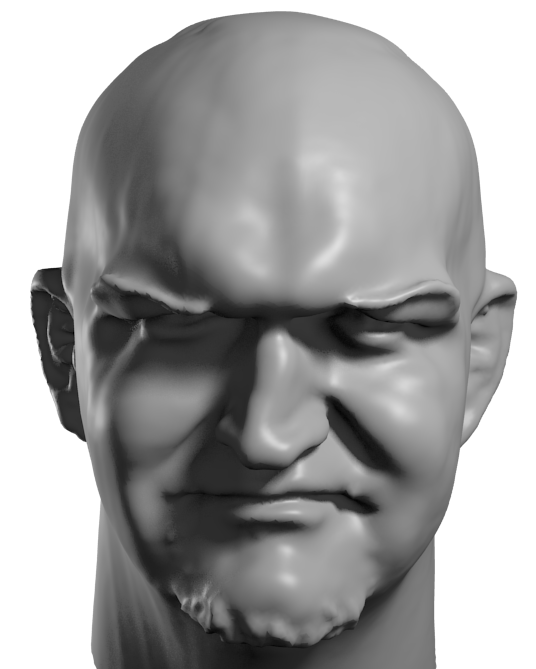}}
\subfloat[]{\includegraphics[width=\imw]{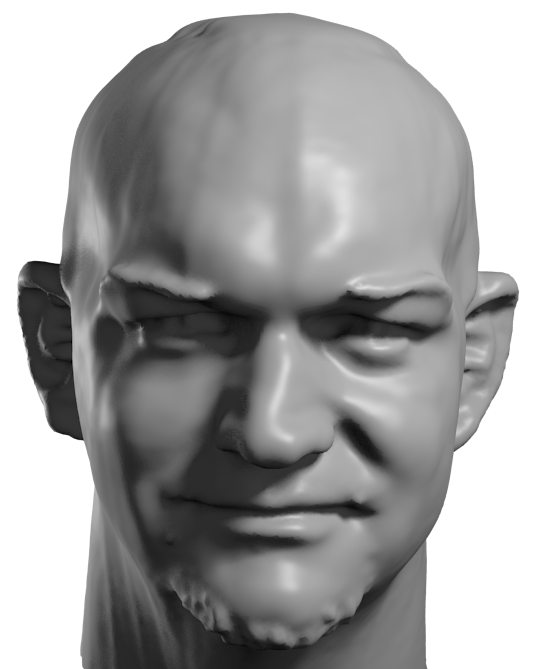}}
\caption{(a) original mesh, (b)-(d) results from our system, (e) result from PriMo \protect \cite{Botsch:2006:PCP}, (f) result using  the screened Poisson equation \protect\cite{Chuang:Kazhdan:SIG2011}. Our system is capable of a wider range of custom stylizations: (b) bulbous nose, sharpened chin and accentuated brow, (c) planarization with low exaggeration, (d) rounder face, larger cheeks. }
\label{fig:primo_comparison}
\end{figure*}

A video and a selection of the results (including original, segmented, abstacted and stylized meshes) are made available on our project webpage:~\url{http://wp.doc.ic.ac.uk/robotvision/project/face-stylization/}

\section{Discussion and Conclusion}
\label{sec:conclusion}
Experts in sculpture will note that both the approach and algorithm we have chosen are both oversimplifications of a
sculpting process. Our interpretation of the planes of the head, inspired by Lanteri and Asaro, is only one abstraction used by sculptors, and our algorithm is only one possible implementation of this abstraction. We hope
and expect that future researchers will explore alternate interpretations and methods with us. We expect that much further insight will be gained by extending the initial comparative study of the geometrical properties of real versus sculpted faces which we present in our supplementary material.

Both Nakpil and Magrath described working at multiple scales in their sculpting work, from large planes to smaller forms. In this paper we have utilized only two scales, but combining deformations at multiple scales may produce a wide variety of sculptural styles.

One significant limitation of our approach is that it works best on surfaces with a mixture of convex and concave regions. (As a trivial example, it is impossible to uniformly decrease all the angles of a convex polygon, because the sum of the internal angles is constant.) Further, our method is largely \emph{ad hoc}, based on conversations with sculptors and our analytical interpretation of their intent. But we believe that given an appropriate dataset, it may be possible to learn sculptural abstraction in a more principled way from real data.

This paper describes just one small step in the direction of sculptural abstraction. We look forward to further exploring the ways in which sculptors depict natural shapes, and learning effective geometric modifications that emulate these techniques.

\section*{Acknowledgments}
\label{sec:ack}
The authors wish to thank Gio Nakpil and Mike Magrath for their insight into sculpting technique. Additional thanks goes to Gio for meeting with us and allowing us to screencast his workflow when modifying a scanned bust, seen in the video, and to Jaime Labelle at Pixologic for providing an evaluation copy of ZBrush.

We thank David Salesin, Holger Winnem\"oller, Wil Li, and Aaron Hertzmann for their input on non-photo-realistic rendering for 3D.

\begin{table*}[!t]
\begin{center}
	\begin{tabular}{ | c | p{10cm} | p{5cm} | }
		\hline
		Page & Passage & Constraint \\ \hline
		15 & ``...with your calipers measure the distance from corner to corner of the mouth, and set this distance off on your horizontal line.'' & distance between mouth corners\\ \hline
		19 & ``measure with calipers the length of the nose'' & distance from bridge to tip of nose \\ \hline
		43--44 & ``calculate or measure (from the place where you fixed the articulation of sternum and collar-bone) the height of the chin...'' & absolute position of chin tip \\ \hline
		48 & ``Measure on your model the distance from the notch of the ear to the most projecting part of the nose-tip (Fig. 38) (take this measure on both sides, for you will frequently find that the distance on right and left side varies)'' & absolute position of nose tip \\ \hline
		50 & ``...the distance from the chin to the eyebrows... by describing an arc on the model, in order to find out if the eyebrows are of the same height on either side...'' & distance from chin to each eyebrow (2) \\ \hline
		53 & ``...model the upper jaw, marking at once the two corners of the mouth, well observing their relation to the size of the nostrils.'' & distance from each corner of mouth to same side of nose (2) \\ \hline
		58--59 & ``A careful measurement from inner corner to inner corner...'' & distance between inner corners of eyes \\ \hline
	\end{tabular}
\end{center}
\caption{\label{tab:Lanteri}Facial feature constraints and their motivating passages from Lanteri [1985].}
\end{table*}

\bibliographystyle{acmsiggraph}
\bibliography{paper}

\begin{appendices}
\section{-\,\,\, Lanteri Constraints}
\label{app:Lanteri}
In selecting geometric constraints to preserve individual facial characteristics, we identified seven passages of Lanteri~[1911, 1985] that correspond to nine relative measurements between points on the surface of the model. All of these passages refer to maintaining distances between points, but our optimization only modifies the face region, so measurements to points such as the ears or the top of the sternum are treated as absolute positional constraints rather than relative ones. Table~\ref{tab:Lanteri} provides these seven passages, along with our interpretation of each as a distance constraint.

\end{appendices}

\end{document}